\documentclass[%
 reprint,
 amsmath,amssymb,
 aps, nofootinbib, superscriptaddress
]{revtex4-2}

\usepackage{graphicx}
\usepackage{amsmath}
\usepackage{physics}
\usepackage{bm}
\usepackage[dvipsnames, usenames]{xcolor}
\usepackage[normalem]{ulem}
\usepackage{appendix}
\usepackage{soul}
\usepackage{nameref}
\usepackage{enumitem}

\makeatletter
\newcommand\setcurrentname[1]{\def\@currentlabelname{#1}}
\makeatother

\newcommand{\appropto}{\mathrel{\vcenter{
  \offinterlineskip\halign{\hfil$##$\cr
    \propto\cr\noalign{\kern2pt}\sim\cr\noalign{\kern-2pt}}}}}

\begin{document}

\title{Axion String Source Modelling}

\author{Amelia Drew}
 \email{a.drew@damtp.cam.ac.uk}
 \affiliation{%
 Centre for Theoretical Cosmology, Department of Applied Mathematics and Theoretical Physics,
University of Cambridge, Wilberforce Road, Cambridge, CB3 0WA, United Kingdom
}%
 \affiliation{Homerton College, Hills Road, Cambridge, CB2 8PH, United Kingdom}
\author{Tomasz Kinowski}%
 \email{tk593@cam.ac.uk}
 \affiliation{Gonville and Caius College, Trinity Street, Cambridge, CB2 1TA, United Kingdom}
 \author{E. P. S. Shellard}%
 \email{e.p.s.shellard@damtp.cam.ac.uk}
 \affiliation{%
 Centre for Theoretical Cosmology, Department of Applied Mathematics and Theoretical Physics,
University of Cambridge, Wilberforce Road, Cambridge, CB3 0WA, United Kingdom
}%

\date{\today}

\begin{abstract}

In this paper, we investigate the effect of the string radius of curvature $R_\mathrm{\,Gaussian}$ on the massive and massless scalar radiation emitted from collisions of travelling waves propagating along an axion (global cosmic) string. We construct initial conditions for two colliding Gaussians and perform parameter scans over their amplitude $A$ and standard deviation $\sigma_\mathrm{d}$. We show that these collisions emit isotropic bursts of massless radiation and that the energy emitted via this channel obeys a power law $\appropto A^{\gamma}$, where the coefficient $\gamma$ depends on the regime of $A/\delta$ and $\delta$ is the string width.  Massive radiation is exponentially suppressed $\appropto e^{-\zeta R_\mathrm{\,Gaussian}}$ in the quasi-linear regime $\sigma_\mathrm{d} \gg \delta$ and exhibits power-law decay $\appropto R_\mathrm{\,Gaussian}^{-\gamma'}$ in the nonlinear regime $\sigma_\mathrm{d} \lesssim 2\delta$, with different $\zeta$ and $\gamma'$ in different regimes of $R_\mathrm{\,Gaussian}$. In this nonlinear regime, massive particle radiation can comprise up to 50\% of the total energy emitted. Drawing on a known parallel between axion radiation from global strings and gravitational radiation from Abelian-Higgs strings, this suggests that massive particle radiation may become significant with respect to the massless (gravitational) channel for such nonlinear burst signals. For all configurations studied, we obtain a spectral index $q \gtrsim 1$ for the axion radiation, where $q \rightarrow 1$ as $A$ increases, i.e. a higher proportion of radiation is emitted in high frequency modes.

\end{abstract}

\maketitle

\section{Introduction}

Axion strings are a class of cosmic string which arise from the breaking of a global $U(1)$ symmetry. They are motivated by simple extensions to the Standard Model, such as certain axion dark matter models, grand unified theories (GUTs) and string theory, and arise cosmologically as a result of a `phase transition' in the early Universe. Neglecting any coupling to gravity, axion strings will radiate energy via massless axion radiation and massive particle radiation. The mechanism behind these decay channels is a crucial point of discussion for two fields within cosmology and astrophysics: axion string network evolution and gravitational wave (GW) modelling. For the former, the spectrum of the massless radiation emitted from a network has direct implications for the predicted mass of the axion in the post-inflationary symmetry breaking scenario \cite{Buschmann_2022, Benabou2023, Gorghetto:2021, Hindmarsh2021, Kawasaki:2018bzv, OHare2022, Marsh:2015xka}. For the latter, the size distribution of cosmic string loops and their decay channels determines the gravitational waveform \cite{Vilenkin:2000jqa, Siemens2006, Blanco-Pillado2013, Lorenz2010, Ringeval2007}. This is crucial both for accurate matched template searches by LIGO-Virgo-KAGRA \cite{Abbott2018, Abbott2021} and similar future GW experiments, and for stochastic background searches, such as the pulsar timing array experiments European Pulsar Timing Array (EPTA) \cite{EPTA} and NanoGRAV \cite{Nanograv15}.

Typically, cosmic string simulations are undertaken either by approximating the string as being infinitely thin using the Nambu-Goto action, or simulating the fundamental fields themselves, known as the `field theory' approach. Radiation in the Nambu-Goto model is well-understood, and gravitational waveforms (see e.g. \cite{Vilenkin:2000jqa, Damour:2000wa, Damour2001, Damour2005}) and power spectra \cite{Sakellariadou1990, Sakellariadou:1991sd, Battye1993} have been calculated for several string configurations. However, these predictions do not have the benefit of incorporating radiation backreaction.

Numerical investigations have also been undertaken into the radiation from individual string configurations in field theory. These do incorporate backreaction, albeit with limited dynamic range. These have included the decay \cite{Blanco-Pillado_2023, Saurabh_2020, Matsunami_2019, baezaballesteros2023} and gravitational collapse \cite{Aurrekoetxea_2020, Aurrekoetxea_2022} of individual loops, radiation from standing waves \cite{Olum2000, Drew2019, Drew2023} and radiation from cusps on Abelian-Higgs strings \cite{Olum1999}. It has been shown that cosmic string loops \cite{Blanco-Pillado_2023} and domain walls \cite{Blanco-Pillado:2022rad} deviate from Nambu-Goto-like behaviour in regions of high string curvature , where curvature can be taken as a proxy for high acceleration. Recent work \cite{Hindmarsh_2023} has also parameterised the proportion of energy that is emitted via particle radiation compared to gravitational radiation for Abelian-Higgs network simulations. A natural question to ask might be: can we extend any properties of the radiation from these specific configurations to more general string configurations, such that we do not need to simulate each one individually?

In this paper, we present investigations of the massive and massless (axion) radiation from axion string network simulations, and from individual cusp-like configurations on an axion string using high resolution adaptive mesh refinement techniques. Section \ref{AxionStringTheory} outlines the theory of global strings, the diagnostics used to analyse their radiation and how to interpret the simulation results. Section \ref{Networks} presents qualitative observations from axion string network simulations, along with their numerical implementation. Section \ref{BurstSignal} details the initial conditions and numerical implementation of cusp-like configurations, obtained by colliding two Gaussian travelling waves along the string. In Section \ref{massiveaxionradiation}, we present results from a detailed parameter scan over the string curvature radius $R_\mathrm{\,Gaussian}$ for the burst configurations, and Section \ref{Discussion} compares the results obtained with previous investigations of sinusoidal configurations \cite{Drew2019, Drew2023}. Section \ref{spectrumdiscussion} discusses the potential implications of these results for GW signals from Abelian-Higgs strings, motivated by \cite{Battye1993}. Finally, we conclude and discuss the implications of this work in Section \ref{Conclusion}. We use `natural' units throughout, setting $\hbar = c = k_B = 1$ such that $[E] = [M] = [L]^{-1} = [T]^{-1}$.

\section{Axion String Theory and Radiation Diagnostics}\label{AxionStringTheory}

In this section, we provide a brief outline of the model for global cosmic strings and the radiation diagnostics used in this paper. Further information can be found in \cite{Vilenkin:2000jqa} and \cite{Drew2019}. 

We consider the Goldstone model for a single complex scalar field $\varphi$ with a symmetry-breaking potential. This has a Lagrangian density $\mathcal{L}$ given by
\begin{equation}
    \mathcal{L} = (\partial_\mu\bar{\varphi})(\partial^\mu\varphi) - V(\varphi)\,,
    \label{GoldstoneLagrangian}
\end{equation}
with the potential
\begin{equation}\label{potential}
V(\varphi) = \frac{1}{4}\lambda(\bar{\varphi}\varphi - \eta^2)^2\,.
\end{equation}
The constant $\eta$ sets the symmetry breaking scale, and the mass of the Higgs particle in the broken symmetry state $m_H$ is set by $\eta$ and $\lambda$, where $m_{\rm H} = \sqrt{\lambda} \,\eta$.

The Euler-Lagrange equations for the evolution if the scalar field are given by 
\begin{equation}
    \frac{\partial^2\phi_{1,2}}{\partial t^2} -
    \nabla^2\phi_{1,2} + \frac{\lambda}{2}\phi_{1,2}(|\varphi|^2 - \eta^2) = 0\,,
     \label{EL}
\end{equation}
where the complex field $\varphi$ is split into its real and imaginary parts by $\varphi = \phi_1 + i\phi_2$. We can numerically solve the static Euler-Lagrange equations,
\begin{equation}\label{initialprofile}
    \pdv[2]{\phi}{r}+\frac{1}{r}\pdv{\phi}{r} -\frac{1}{r^2}\phi - \frac{1}{2}\phi(\phi^2-\eta^2) = 0 \,,
\end{equation}
in cylindrical symmetry and subject to the boundary conditions $\phi(0) = 0$ and $\phi \rightarrow \eta$ as $r \rightarrow \infty$, to obtain the radial string profile $\phi(r)$. The initial conditions for a string with winding number $n=1$ are then given by $\varphi(r,\theta) = \phi(r)e^{i\theta}$, where $\phi = |\varphi|$ and $\theta$ is the angle in cylindrical polar coordinates. For the string network configurations in this paper, we utilise a period of dissipative evolution to form the network prior to the Euler-Lagrange evolution \eqref{EL}, using
\begin{equation}
    \frac{\partial\phi_{1,2}}{\partial t} -
    \nabla^2\phi_{1,2} + \frac{\lambda}{2}\phi_{1,2}(|\varphi|^2 - \eta^2) = 0\,.
     \label{dissipativeevolution}
\end{equation}

\subsection{Radiation and Energy Diagnostics}

The energy density of $\varphi$ can be split into massive and massless components using the $T^{00}$ component of the stress-energy tensor $T^{\mu\nu}$ in the following form \cite{Drew2019, Drew2023}:
 \begin{eqnarray}\label{energymomdiag}
   T^{00} =  {\Pi}_\phi^2 + ( {\cal D} \phi)^2 + {\Pi}_\vartheta^2 + ( {\cal D} \vartheta)^2 +  {\textstyle\frac{\lambda}{4}} (\phi^2 -\eta^2)^2\,.
\end{eqnarray}
The massive components are given by
\begin{alignat}{4}\label{massivediagnostic}
 &{\Pi}_\phi& &\equiv{}& &{}\dot{\phi}{}& &= \frac{\phi_1\dot\phi_1 + \phi_2\dot\phi_2}{\phi}, \\ 
 &{\cal D}_i\phi& &\equiv{}& &{}\nabla_i\phi{}& &= \frac{\phi_1\nabla_i\phi_1 + \phi_2\nabla_i\phi_2}{\phi}\,,
\end{alignat} 
and the massless by
\begin{alignat}{4}\label{masslessdiagnostic}
&{\Pi}_\vartheta& &\equiv{}& &{}\phi\,\dot{\vartheta}{}& &= \frac{\phi_1\dot\phi_2 - \phi_2\dot\phi_1}{\phi}, \\ 
&{\cal D}_i\vartheta& &\equiv{}& &{}\phi\,\nabla_i\vartheta{}& &= \frac{\phi_1\nabla_i\phi_2 - \phi_2\nabla_i\phi_1}{\phi}\,,
\end{alignat}
where the potential contribution tends to zero at large distances from the string core, and $\phi$ and $\theta$ have been promoted to a dynamical variables, $\phi = \phi(x^\mu)$ and $\vartheta = \vartheta(x^\mu)$. The tension of the string itself is calculated by integration of $T^{00}$ over the polar angle and up to an appropriate radius, and is given by
\begin{equation}\label{tension}
    \mu \approx \mu_0 + 2\pi\eta^2 \ln (R/\delta)\,,
\end{equation}
where $\mu_0$ is the contribution from the massive string core and the second term is the contribution from the long-range massless field, which dominates at $R \gtrsim 2\delta$. Here, $\delta = m_H^{-1}$ is the string width and $R$ is the approximate curvature scale of the string, i.e. the radius out to which it is appropriate to integrate.

The momentum component $T^{0i}$ of the stress tensor can also be split into components, given by
\begin{eqnarray}\label{momentumdiag}
   P_i\equiv T^{0i} =  2({\Pi}_\phi {\cal D}_i \phi + {\Pi}_\vartheta{\cal D}_i \vartheta) \,,
\end{eqnarray}
where the two terms represent the massive and massless radiation energy fluxes respectively. We can calculate the magnitude of these two components that are propagating away from a string in the radial direction ${\bf P}\cdot \hat{\bf r}$ and integrate over a surface $S$ at large $r$ over time $t$ to determine the total energy flux through the surface for each channel. The massive component is given by
\begin{align}\label{Pmassive}
E_{\mathrm{massive}} &= \int P_{\mathrm{massive}} \, \mathrm{d}t \\ \nonumber &\propto \int{(\Pi_{\phi}\mathcal{D}\phi)\cdot\mathbf{\hat{r}}\,\mathrm{d}S \mathrm{d}t}\,,
\end{align}
and the massless component by
\begin{align}\label{Pmassless}
E_{\mathrm{massless}} &= \int P_{\mathrm{massless}} \, \mathrm{d}t \\ \nonumber &\propto \int{(\Pi_{\vartheta}\mathcal{D}\vartheta)\cdot\mathbf{\hat{r}}\,\mathrm{d}S \mathrm{d}t}\,.
\end{align}
These diagnostics are used for the majority of the analysis in this study. 

\subsection{Parameters and Interpretation of Simulations}

The parameters $\eta$ and $\lambda$ that determine the symmetry breaking scale via the potential \eqref{potential} can both be scaled out of the Euler-Lagrange equations \eqref{EL} without loss of generality. This is achieved by rescaling $\phi \rightarrow \phi/\eta$, $x \rightarrow \eta\,x$ and $t \rightarrow \eta\,t$ allowing us to set $\eta = 1$, and rescaling $x \rightarrow \sqrt{\lambda}\,x$ and $t \rightarrow \sqrt{\lambda}\,t$ allowing us to scale out $\lambda$. We choose $\lambda=3$ for the network simulation in Section \ref{Networks} and $\lambda=1$ for the travelling wave simulations. This means that numerical results from simulations of a given string configuration can be rescaled to different energies, with the ratio of string curvature $R$ to string width $\delta$ as a free parameter. For example, changing $\lambda$ for a curved string is equivalent to changing $R/\delta$, and vice versa. 

Setting $\eta = 1$ sets the units of time, each length dimension and the value of $|\phi|$. In order to interpret results from our simulations in a specific context, we can reintroduce $\eta$ as necessary. For example, for the measured energy density $T^{00}$, we multiply by $\eta^4$ to obtain the rescaled value. From any string configuration we simulate, the ratio that we observe between the massive and massless radiation is therefore independent of $\eta$; even if the overall energy scale of the simulated configuration changes, the proportion of massive and massless radiation emitted does not. For configurations in which we see massless and massive radiation approximately equivalent in magnitude, this will also be the case in reality.

Physically, the role of $\lambda$ is to set the mass threshold $m_H \sim \sqrt{\lambda}\eta$ which must be overcome for massive radiation to be able to propagate, as shown in \cite{Drew2023}. This means that the energy scale $\eta$ does not fully determine the magnitude of the massive decay channel, and $\lambda$ also has an effect via this mechanism. We will see that the details of this also depend on the configuration of the source, specifically the ratio $R/\delta \approx R\sqrt{\lambda}\eta$.

\section{Network Simulations}\label{Networks}

\begin{figure*}
    \centering        \includegraphics[width=0.48\textwidth, trim=120 30 0 280, clip]{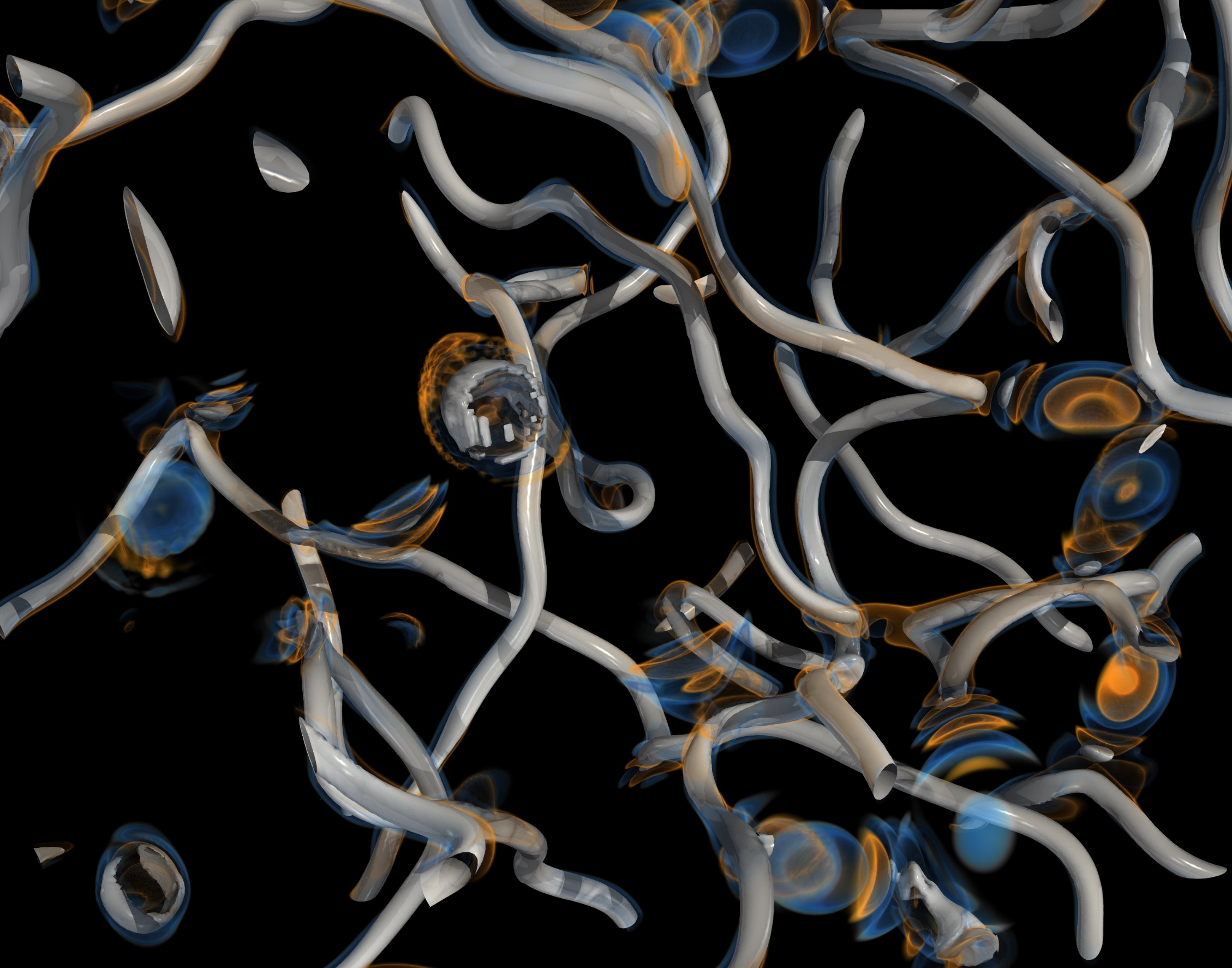}
    \includegraphics[width=0.48\textwidth, trim=120 30 0 280, clip]{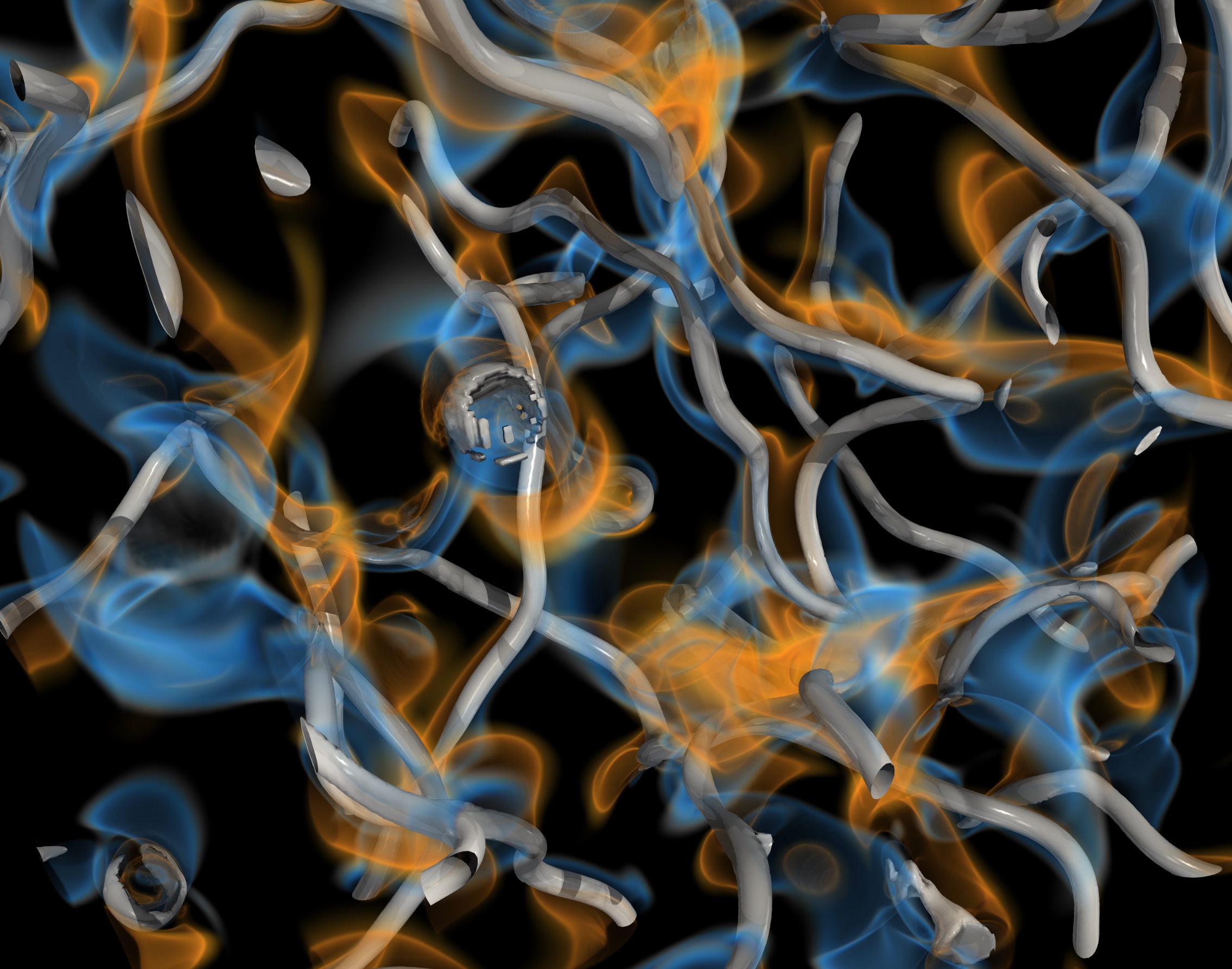} \\
    \includegraphics[width=0.48\textwidth, trim=0 12 0 15, clip]{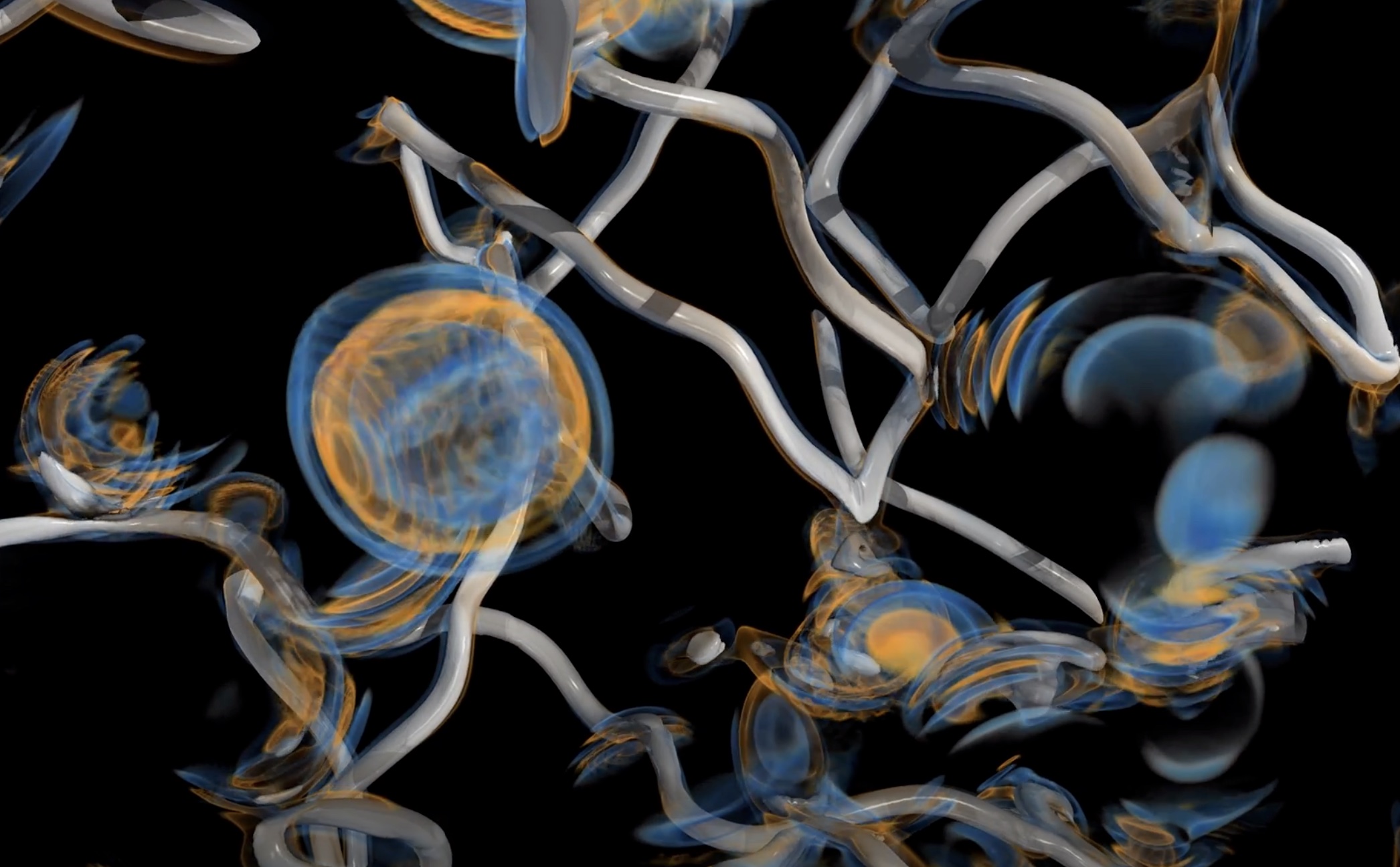}
    \includegraphics[width=0.48\textwidth, trim=60 0 60 0, clip]{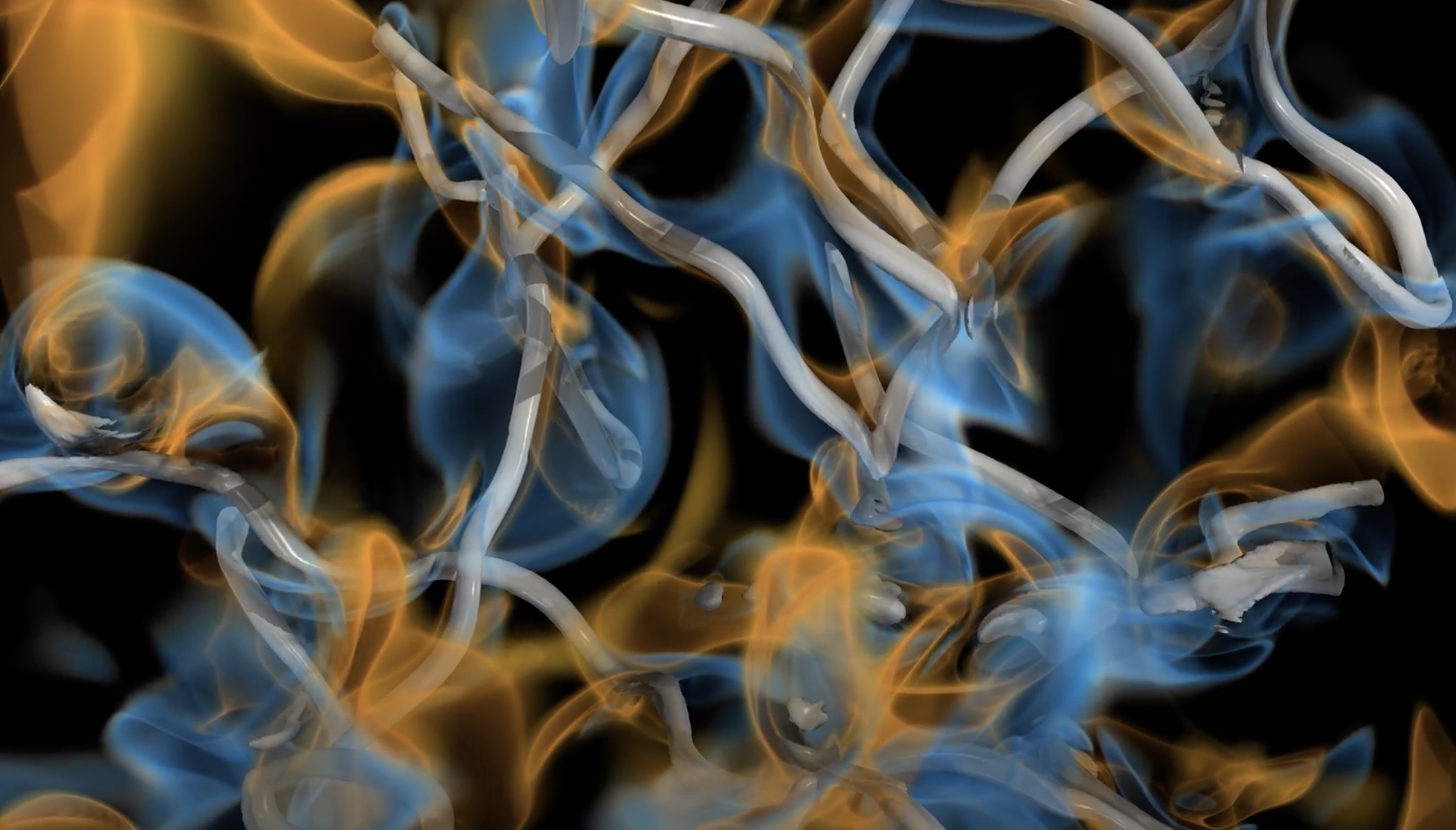} \\
    \includegraphics[width=0.48\textwidth, trim=0 100 0 160, clip]{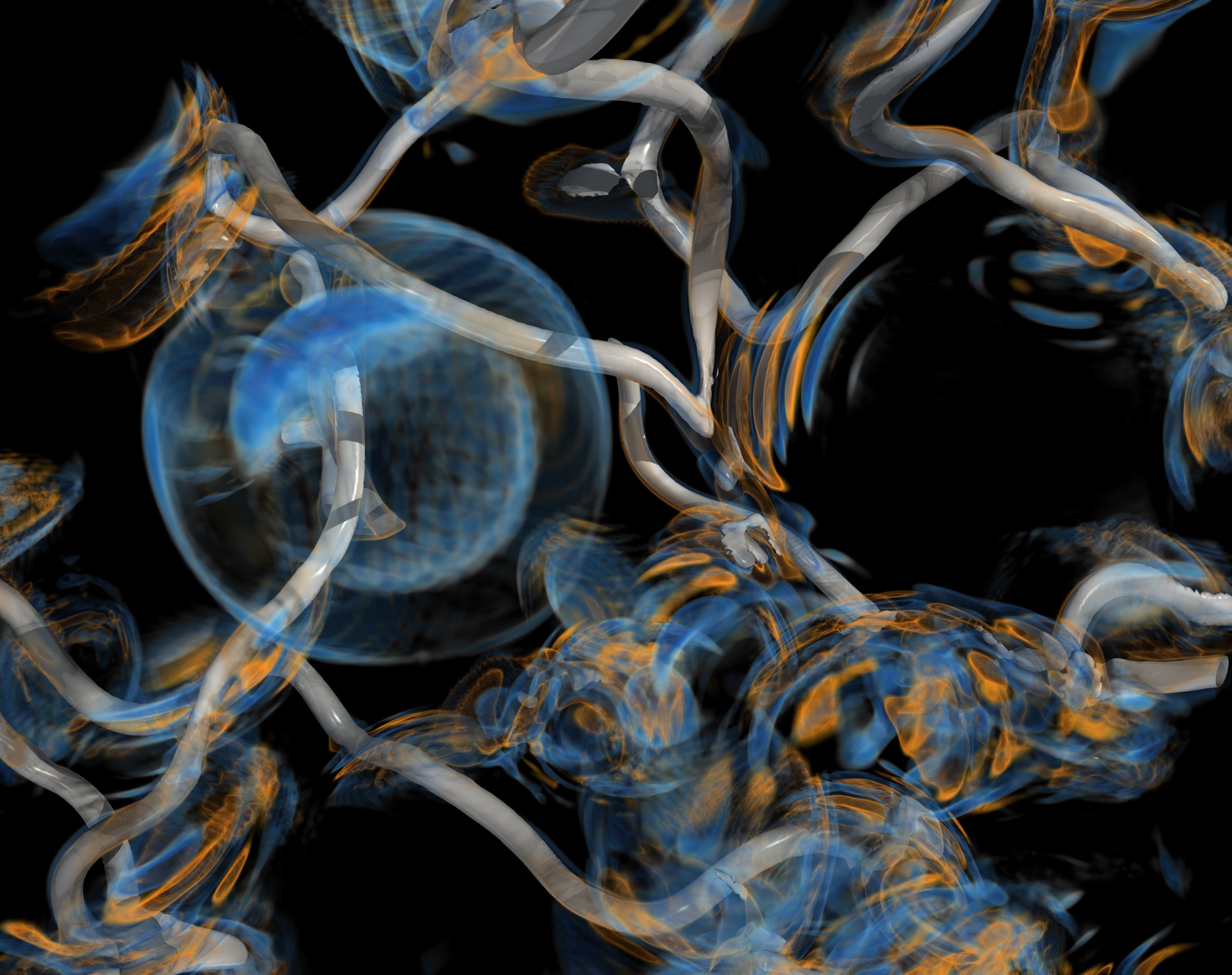}
    \includegraphics[width=0.48\textwidth, trim=0 100 0 170, clip]{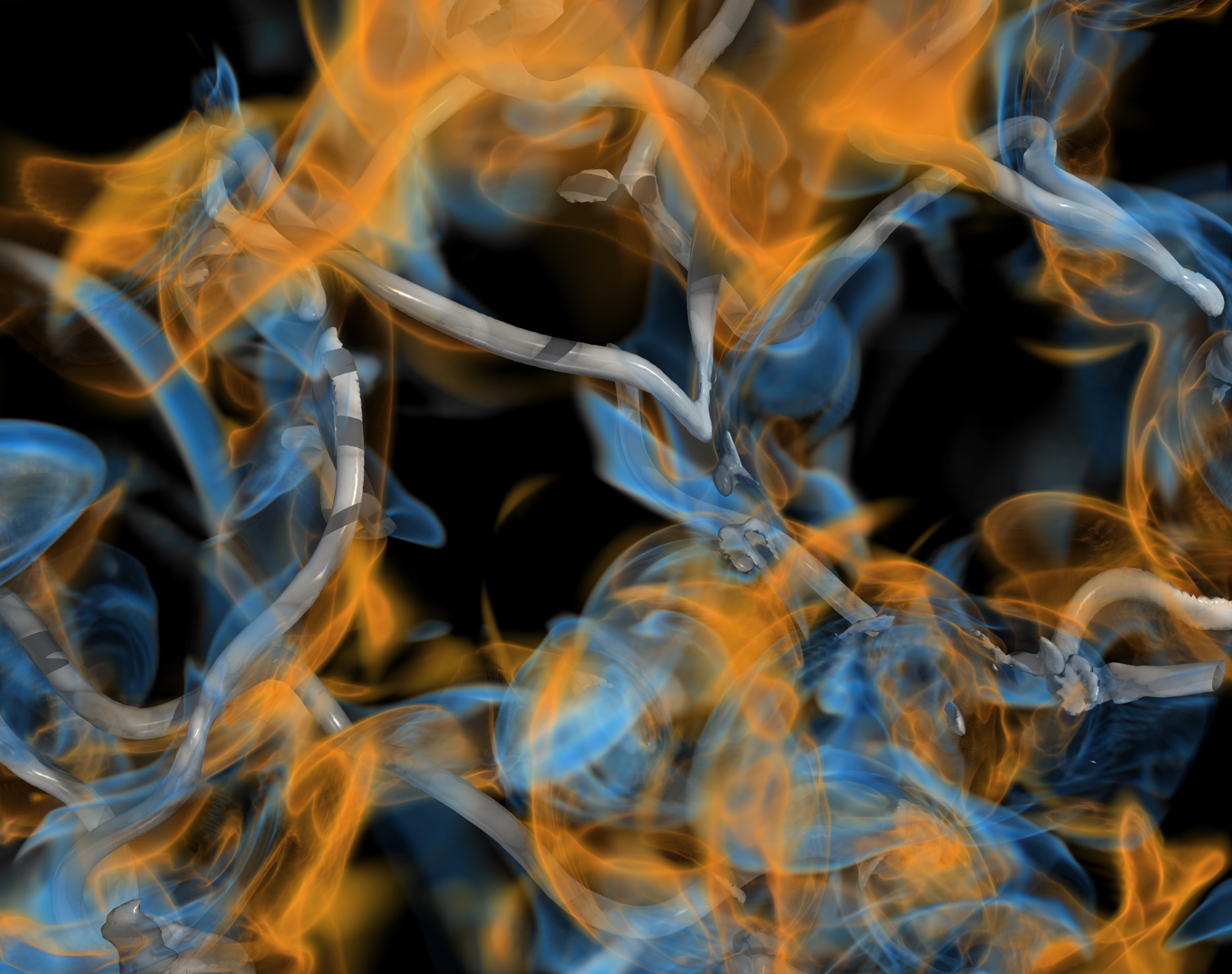}
    \includegraphics[width=0.48\textwidth, trim=0 100 0 100, clip]{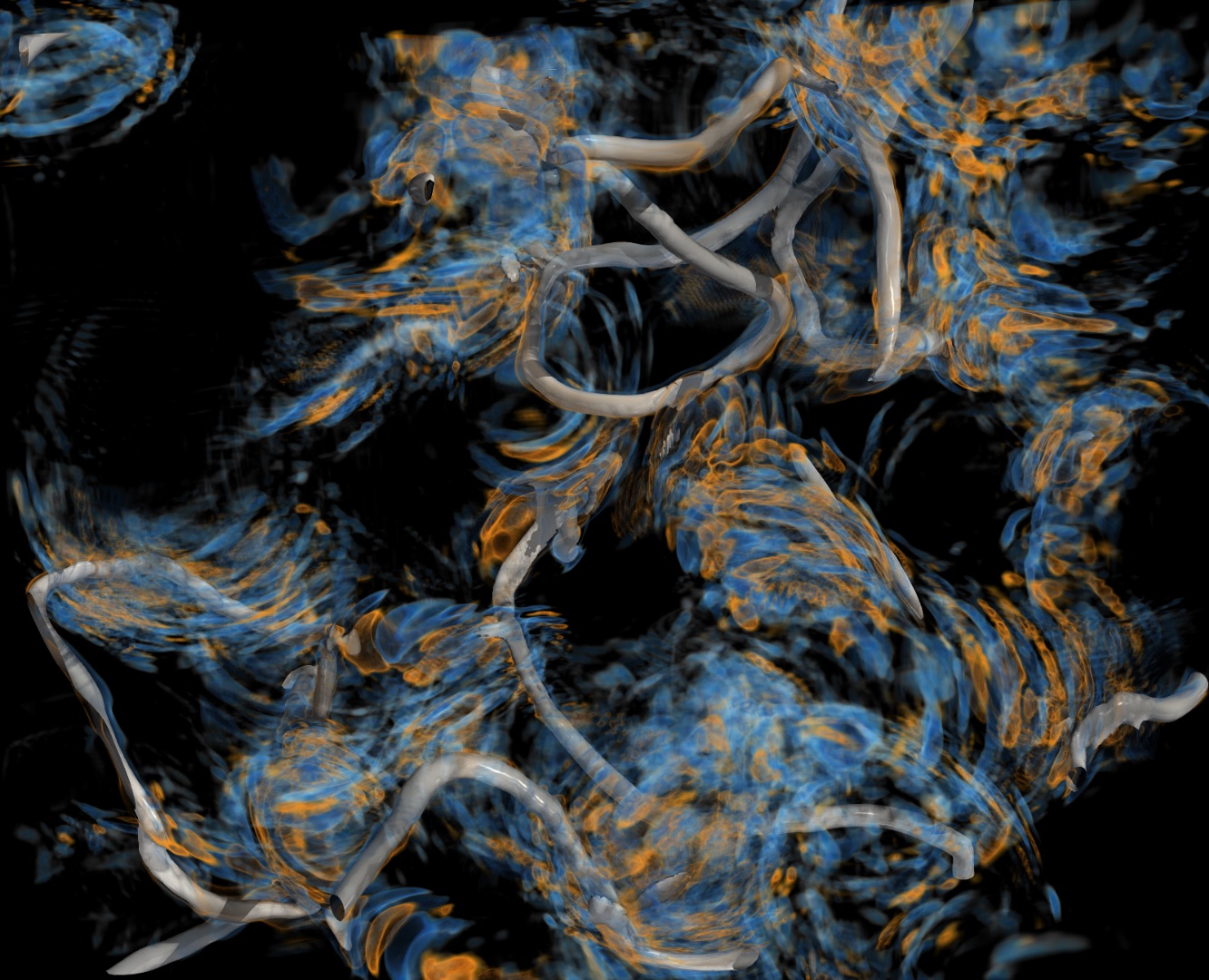}
    \includegraphics[width=0.48\textwidth, trim=0 100 0 100, clip]{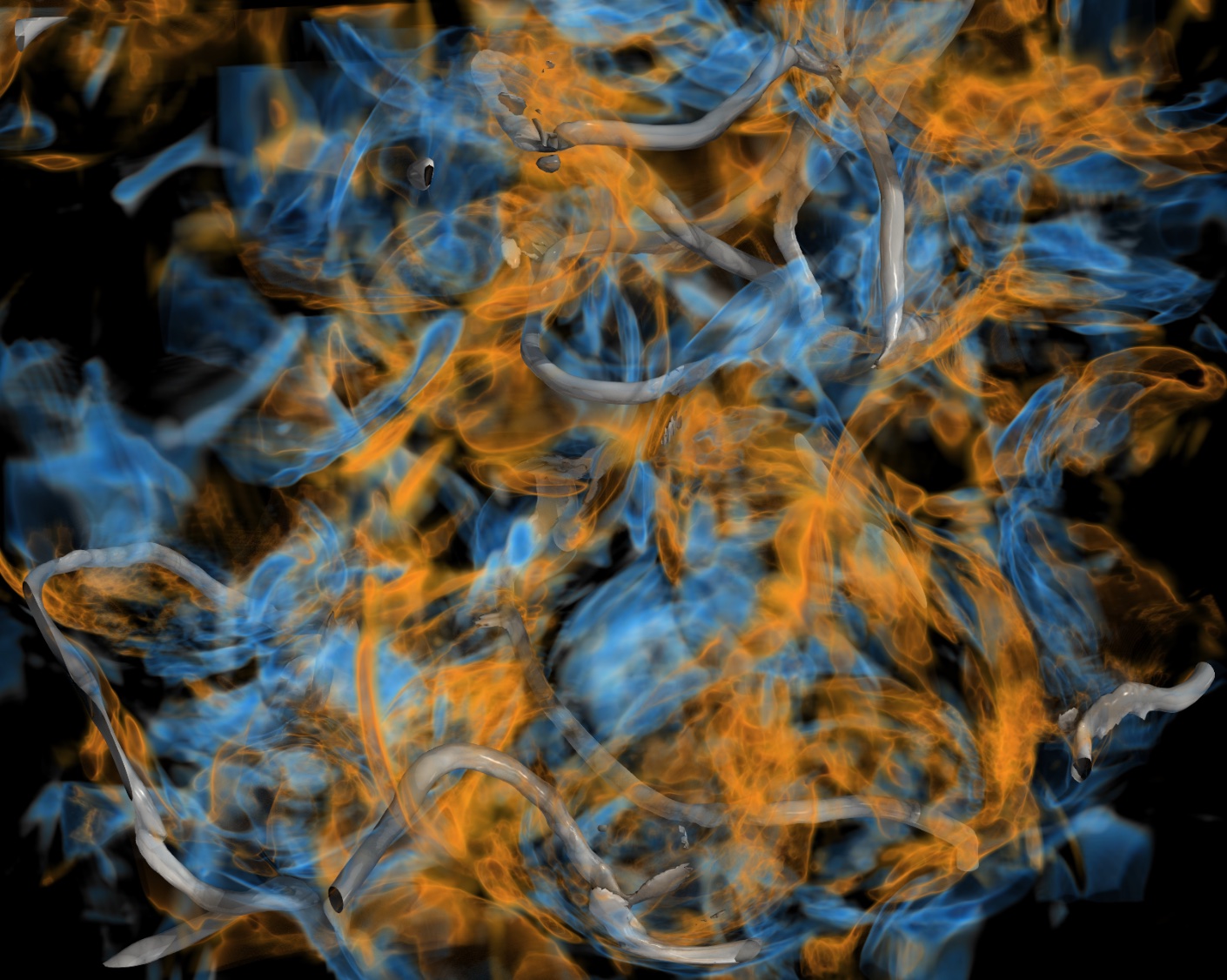}    
    \caption{Volume rendering in 3D space $(x,y,z)$ of massive ${\Pi}_\phi$ (left) and massless ${\Pi}_\vartheta$ (right) radiation from a $\lambda=3$ string network at $t\approx 15,\, 20,\, 30$ and $50$ from top to bottom. Strings are indicated by grey contours around the cores, and both channels of radiation are indicated in blue and yellow (maxima and minima).}
    \label{networkradiation2}
\end{figure*}

In this section, we present qualitative observations from an adaptive mesh simulation of a global cosmic string network in a flat background. The simulation is carried out using $\texttt{GRChombo}$ on a $128^3$ base grid, with periodic boundary conditions, string width $\lambda=3$ and initial conditions obtained as described in Section \ref{NetworkICs}. We use a regridding threshold $|\phi_{\mathrm{threshold}}|=0.5$, a coarsest grid resolution $\Delta x_0 =1$, coarsest timestep $\Delta t_0 = \Delta x_0 /100$ and Kreiss-Oliger damping coefficient of 0.3. The simulation uses a maximum refinement level of $l_\mathrm{max}=4$, with $\Delta x = 0.0625$.

\subsection{Network Initial Conditions}\label{NetworkICs}

The initial conditions for a string network are obtained numerically by assigning a random phase $-\pi \leq \theta < \pi$ and magnitude $-0.01 \leq \phi \leq 0.01$ to each gridpoint on the coarsest level of the grid. We assign
\begin{equation}
    \phi_1 = \phi\cos{\theta}, \hspace{1cm} \phi_2 = \phi\sin{\theta}\,.
\end{equation}
This simulates the complex scalar field $\varphi$ falling into the potential minimum of $V(\varphi)$ at different values of $\theta$. This initial configuration can be evolved using the dissipative evolution equations (\ref{dissipativeevolution}) until a network of distinct strings has formed. We identify the formation of strings by visualising $\phi$ in three dimensions at various stages of damping, using \texttt{Paraview} to determine when regions between the strings reach minimal energy and the string width has stabilised, i.e.\ when high frequency internal degrees of freedom have been damped.\footnote{In a cosmological scenario, the mechanism for dissipation comes from the expansion of the Universe.} We subsequently evolve using the wave equations \eqref{EL}. These initial conditions have proved useful for qualitative observations of string radiation.

\subsection{Qualitative Evolution}

Figure \ref{networkradiation2} shows volume renderings of the massive diagnostic ${\Pi}_\phi$ \eqref{massivediagnostic} and massless diagnostic $\Pi_\vartheta$ \eqref{masslessdiagnostic} for radiation emitted from the network at $t\approx 15,\, 20,\, 30$ and $50$. The massless diagnostic $\Pi_\vartheta$ includes contributions from the string self-field, which we bear in mind when drawing qualitative conclusions. At $t \approx 15$ we observe from the massive radiation that directed bursts are beginning to be emitted from nonlinear configurations formed as a result of string reconnections, as well as signals from collapsing loops. The massless radiation emitted is more diffuse, and is becoming increasingly significant. At $t \approx 20$, the visualisation shows clear beamed massive signals from relativistic sections and approximately spherical blast waves from loop collapse, with the massless radiation still relatively diffuse. We also observe that the string network density is beginning to decrease as loops collapse and strings are annihilated. Both of these trends continue at $t \approx 30$, and finally at $t \approx 50$ we observe more diffuse massive and massless radiation distributed throughout the simulation box, again with a decreased density of strings. In general, we observe that massless radiation emanating from the strings is spread quite diffusely throughout the volume, whereas massive radiation is more localised to the specific configurations described.

Several interesting qualitative conclusions can be drawn from the contrasting nature of the massive and massless radiation from a string network. Although the massive radiation signals are striking, particularly the dramatic and explosive demise of small loops, we note that all of these massive signals are localised to regions of high curvature on scales comparable to the string width. For regions where the string motion is coherent and the curvature is lower, massive radiation is less evident. In comparison, massless radiation is also emitted from these high curvature regions, but is more pervasive from all strings in the evolving networks. Whereas the simulation box is largely filled with massless radiation at the late stage, there are still voids evident in the massive radiation. This has important implications for the extrapolation to cosmological strings, where the existence of curved regions comparable to the string width likely occurs much less frequently. We may therefore expect massive radiation to be suppressed and localised, especially relative to the global emission of massless modes.

\section{Burst Signal Implementation}\label{BurstSignal}

In this section, we discuss our implementation of travelling wave initial conditions, our numerical grid setup and results from convergence tests. In Section \ref{travellingwave}, we present the analytic equations that describe collisions of travelling waves with a Gaussian profile, which we implement to emulate realistic cusp-like or `burst' configurations. This allows us to maintain direct control over important parameters such as the radius of curvature of the string. Section \ref{NambuGotostrings} details the relationship between these travelling wave solutions and analytic Nambu-Goto string solutions, which are often used in cosmic string modelling. Sections \ref{numericalsetup} and \ref{convergence} detail the numerical setup and convergence test results, including a comparison to fixed grid results.

\subsection{Travelling Wave Initial Conditions}\label{travellingwave}

It has been shown that analytic field theory solutions can be obtained for travelling wave configurations on a global cosmic string which are consistent with Kalb-Ramond modelling \cite{Vachaspati1990}. These are obtained by redefining the coordinate along which direction the string core is displaced, which is transverse to the direction of travel of the wave. We choose to redefine $x \rightarrow X$, where the new $X$-coordinate is given by
\begin{equation}\label{StringCore}
    X = x - \psi(z\,\pm\,t)\,.
\end{equation}
We introduce the function $\psi(z\,\pm\,t)$ to define the shape of the travelling wave and to indicate the direction of travel of the specified configuration along the string, in the positive or negative $z$-direction.  We will comment shortly on the interpretation of these solutions in terms of Nambu-Goto string solutions once transverse degrees of freedom are integrated out.  At this point it is sufficient to note that the field deformations \eqref{StringCore} are considerably more general, allowing the creation of large amplitude, ultrarelativistic configurations (essentially field `shock waves') for which the Nambu-Goto string analogue is not straightforward. Nevertheless, we will study the full-range of possibilities whether or not there are realistic cosmological production mechanisms in some regimes. 

We choose to investigate the collision of two Gaussian configurations, which gives us control over the radius of curvature along the string. We set 
\begin{equation}\label{Gaussian}
    \psi(z \pm t) = \psi_\mathrm{G}(z + t) \pm \psi_\mathrm{G}(z - t)\,,
\end{equation}
where the Gaussian configuration $\psi_G$ is given by
\begin{equation}\label{psiG}
    \psi_G(z \pm t) = A \exp\left\{-\frac{(z \pm t \mp b)^2}{2\sigma_{\mathrm d}^2}\right\}\,.
\end{equation}
Here, $A$ is the amplitude, $b$ is the displacement of the centre of the Gaussian from the centre of the simulation box at $z=0$, and $\sigma_{\mathrm d}$ is the standard deviation. The $\pm$ in \eqref{Gaussian} defines whether we add two Gaussians of the same sign (Gaussian/Gaussian configuration) or of opposite signs (anti-Gaussian/Gaussian configuration). Both configurations are investigated in this study. 

In order to obtain the initial conditions, we substitute the numerically-determined complex scalar field profile from \eqref{initialprofile} with 
\begin{equation}
    \varphi(r,\theta) \rightarrow \Phi(X,y)e^{i\Theta}\,.
\end{equation}
The real and imaginary parts $\Phi_1$ and $\Phi_2$ are now defined by
\begin{equation}\label{newphi}
    \Phi_1 = \Phi(X,y)\cos\Theta, \hspace{1cm} \Phi_2 = \Phi(X,y)\sin\Theta\,,
\end{equation}
where $\tan{\Theta} = y/X$. The initial time derivatives $\Pi_{1,2}$ are obtained by differentiating $\Phi_1$ and $\Phi_2$ with respect to $t$, given by
\begin{align}
    \Pi_1 &= \left.\frac{\partial X}{\partial t}\right|_{t=0}\left(\frac{X^2}{R^2}\frac{\partial\Phi}{\partial R} + \Phi\frac{y^2}{R^3}\right)\,, \nonumber \\ \label{newpi}
    \Pi_2 &= \left.\frac{\partial X}{\partial t}\right|_{t=0}\left(\frac{\partial\Phi}{\partial R}\frac{Xy}{R^2} - \Phi \frac{Xy}{R^3}\right)\,, 
\end{align}
where $R^2 = X^2 + y^2$ and
\begin{align}
    \left.\frac{\partial X}{\partial t}\right|_{t=0} = \left.\frac{\partial\psi}{\partial{t}}\right|_{t=0}\,.
\end{align}
As outlined in \cite{Vachaspati1990}, \eqref{newphi} and \eqref{newpi} can be used to set initial conditions for a travelling wave which can be evolved using the wave equations \eqref{EL}.

It is common in numerical simulations of cosmic strings to apply a period of dissipation to the initial conditions, prior to evolving using the wave equation. Dissipation is less necessary here than, for example, with sinusoidal initial conditions, as we are already using a field theory solution which will be close to a true physical travelling wave configuration. We therefore do not apply dissipative evolution to the initial conditions in this case, allowing us to keep better control of the parameter space. 

\subsection{Correspondence with Nambu-Goto Strings}\label{NambuGotostrings}

The Nambu-Goto action for a string, proportional to its worldsheet area, can be derived in the Abelian-Higgs model by integrating out the transverse degrees of freedom around the cylindrical string solution \eqref{initialprofile}. This methodology can also be extended to axion strings \eqref{tension} to describe them using the Kalb-Ramond action (see \cite{Vilenkin:2000jqa} and references therein).   However, these derivations depend on the underlying assumption that perturbations along the string are small, in particular, that the local radius of curvature $R$ is considerably larger than the string width $\delta \approx 1/(\sqrt{\lambda} \eta)$ over which the integration is taking place, i.e.\ $R \gg \delta$.   The general travelling wave deformations \eqref{StringCore} do not need to satisfy this requirement and, in principle, the Gaussian \eqref{Gaussian} configurations being investigated here, can have an arbitrarily large amplitude (and small curvature radius $R$).  For this reason, it is worth clarifying where the parameter range of this study can be expected to correspond to realistic Nambu-Goto strings or cosmic axion strings. 

We can construct Nambu-Goto long string configurations along which left- and right-moving modes travel, using an analogous solution to \eqref{Gaussian} with the $x$- coordinate of the string core given by
\begin{equation}\label{NGx}
x(\sigma,\tau) = {\textstyle \frac{1}{2}} \left [\Psi_{\rm G} (\sigma-\tau) + \Psi_{\rm G} (\sigma+\tau)\right]\,,
\end{equation}
where $\Psi_{\rm G}(\sigma,\tau)$ is given by \eqref{psiG}, $\sigma$ measures the invariant length along the string and $\tau$ is the proper time.  The magnitude of the derivative of the vector left- and right-moving modes is constrained to be unity, so the $z$-coordinate is given by 
\begin{eqnarray}\label{NGz}
z(\sigma,\tau) &= &{\textstyle \frac{1}{2}} \left [\Phi_{\rm G} (\sigma-\tau) + \Phi_{\rm G} (\sigma+\tau)\right]\,, 
\end{eqnarray}
where  
\begin{equation}\label{NGphi}
\Phi_{\rm G} (\sigma) = \int d\sigma \sqrt{1 - \Psi _{\rm G} '(\sigma)^2}\,.
\end{equation}
If the function $\Phi_G$ is to be real, then the amplitude $A$ of the Gaussian in \eqref{psiG} is constrained to satisfy \begin{equation}\label{NGconstraint}
A_{\rm max}\le \sqrt {e} \sigma_{\rm d}\,,
\end{equation} where $e$ is the elementary charge and can be set equal to unity. The magnitude of the velocity along the string is simply given by the time derivatives of \eqref{NGx} and \eqref{NGz}.

\begin{figure}
    \centering
    \bigskip
    \includegraphics[width=0.45\textwidth]{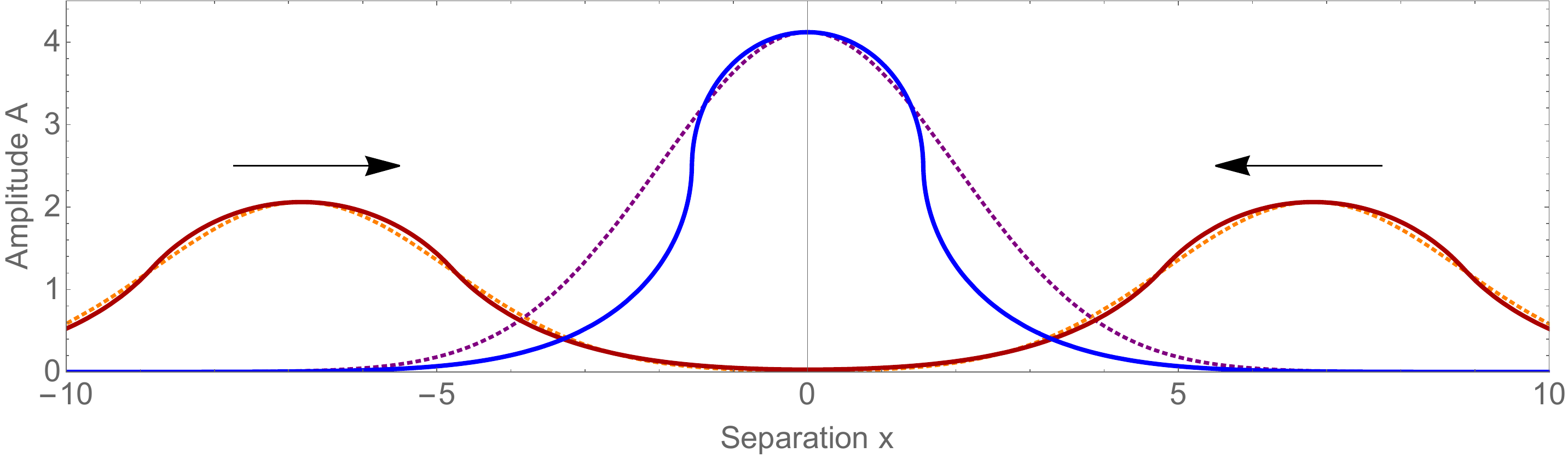}
     \caption{Parametric Nambu-Goto solution \eqref{NGx} with $\sigma_{\rm d,eff} \approx 2.5$ and $A=4.12$ (solid lines) mimicking the dual Gaussian field configuration \eqref{Gaussian} $\sigma_{\rm d} = 2$ and $A=4.12$ (dotted lines). This is the maximal Nambu-Goto solution for this Gaussian width with the string configuration creating `cusps', a point moving momentarily at the speed of light.  Beyond this amplitude $A_{\rm max}$, there are no simple collision Gaussian string solutions. }
    \label{NambuGotosolutions}
\end{figure}

An example of a parametric Nambu-Goto string solution closely corresponding to the `Gaussian burst' configuration \eqref{Gaussian} with $\sigma_{\rm d}=2$ and $A=4$  is shown in Figure \ref{NambuGotosolutions}. This has the maximal amplitude allowed for the Nambu-Goto Gaussian solution, which means that it generates two `cusps' before and after the left- and right-moving modes meet to create the rounded shape at the centre.   A cusp is a single point where the string momentarily attains the speed of light $c$, but in this solution there is also a wider relativistic region around the centre that achieves a large $\gamma$-factor, thus facilitating the generation of a burst of radiation.

How then do we understand the large amplitude travelling wave solutions that violate \eqref{NGconstraint}?   In this case, the field travelling waves do not satisfy the criterion $A\le \sigma_{\rm d}$ and there is not a simple Nambu-Goto solution of the form (\ref{NGx}--\ref{NGz}).  The travelling waves \eqref{Gaussian} can be thought of as highly relativistic field deformations that propagate along straight string, as if `shock waves' or correlated energy packets.  When meeting the counterpart travelling in the opposite direction, there is no simple linear or quasi-linear mechanism by which the two modes superpose and smoothly pass through each other as in the Nambu-Goto case shown in Figure \ref{NambuGotosolutions}.  Instead, at collision a super-relativistic configuration forms that will most naturally annihilate with a huge release of radiative energy (both massless and massive), rapidly reducing the travelling wave towards or below the $A_{\rm max}$ allowed by the Nambu-Goto solution \eqref{NGconstraint}. As we shall see, for extreme amplitudes the string annihilation process can be sufficiently energetic and coherent to form a new loop which reconnects with the original configuration, albeit at greatly reduced amplitude.  It is questionable whether the short-lived travelling wave solutions with $A\gg \sigma_{\rm d}$ created here numerically can also be generated in a realistic physical context, because they require a correlated mechanism by which a large amplitude, highly relativistic string deformation is launched along a straight unperturbed string.

\subsection{Numerical Setup}\label{numericalsetup}

For the parameter scans performed in this paper, it is necessary to balance several numerical and physical factors in order to choose the optimal simulation box configuration. We require a string configuration with an appropriate initial separation of the travelling waves; not too far apart as to add unnecessary time to the simulations, but not too close that the Gaussians overlap at $t=0$. The total string length has to be long enough that the entire signal from the first burst can be extracted before the travelling waves collide with the $z$ boundaries, and the $x$ and $y$ boundaries have to be far enough away from the string to reduce any (very minimal) effects from incoming radiation. On the other hand, the box size has to be small enough for simulations to be performed quickly and using a non-prohibitive number of computational resources; in this case, approximately 200 CPUs. The computational restrictions on the total box size are significantly lessened by the use of adaptive mesh refinement (AMR). 

We find the best configuration to be a total coarse simulation box size of $512^3$ with periodic boundaries in the $z$ direction and Sommerfeld outgoing radiation boundary conditions in the $x$ and $y$ directions. We use an initial Gaussian separation of $d = 2b = 32$ and an extraction cylinder at $R=64$. This means that the travelling waves that collide near the beginning of the simulation collide again on the $z$ boundary at $t\approx300$, so we are able to analyse the simulation confidently up until $t\approx250$ with minimal concerns about the aforementioned sources of numerical error.

All simulations in this paper are performed using the adaptive mesh refinement code, \texttt{GRChombo} \cite{Clough2015, Andrade:2021}. We use a base grid resolution of $\Delta x_0 = 1$ and a base
timestep of $\Delta t_0 = \Delta x_0/4$. Each refinement level reduces $\Delta t$ and $\Delta x$ by a ratio of 2 compared to the next coarsest level. Based on observations from \cite{Drew2023} of the effects of damping high frequency modes, we choose to set the Kreiss-Oliger damping coefficient to be zero. We set the regridding threshold $|\phi_{\mathrm{threshold}}| = 0.25$, prompting additional mesh refinement to be implemented around the string core. The maximum refinement level obtained is $l_\mathrm{max} = 6$ ($\Delta x = 0.015625$) for very high amplitude runs.

\subsection{Convergence Testing}\label{convergence}

We perform convergence tests on Gaussian/Gaussian configurations with the parameters $A=35$, $A=8$ and $A=1$ with $\sigma_{\mathrm d}=2$ in order to investigate the full range of amplitudes run in Section \ref{massiveaxionradiation}, concentrating on configurations in the nonlinear regime. We perform AMR tests using the parameters in Table \ref{convergence_params}, presented in \nameref{AppendixB}, and compare to fixed grid results. We focus on the convergence of $E_\mathrm{massive}$ \eqref{Pmassive} and $E_\mathrm{massless}$ \eqref{Pmassless}. As \texttt{GRChombo} uses a fourth order Runge-Kutta evolution scheme, we expect to see approximately fourth-order convergence. We use these convergence tests, along with any differences between the AMR and fixed grid simulations, to estimate error bars for Figure \ref{fig:AGGRadiation}. Convergence test plots are presented in \nameref{AppendixB}. The comparison with fixed grid results must be taken in context; running a fixed grid simulation, for example, with $\Delta x = 0.0625$ in order to compare with a maximum AMR refinement level $l_\mathrm{max}=4$ is far too computationally intensive, a demonstration in itself of the necessity of AMR. We therefore compare our AMR results to fixed grid results with a coarser refinement $\Delta x = 0.25$ as an \textit{indicator} of accuracy only, bearing in mind that this refinement may not be sufficient for an accurate fixed grid result. 

The AMR convergence test for $A=35$ in Figure \ref{fig:AMRamp35convergence} shows that $E_\mathrm{massive}$ and $E_\mathrm{massless}$ converge at approximately second and third order respectively, although the convergence order for $E_\mathrm{massless}$ varies over time. The AMR simulations reach a maximum refinement level of $l_\mathrm{max}=6$ with $\Delta x = 0.015625$, and $E_\mathrm{massive}$ and $E_\mathrm{massless}$ measured by the finest fixed grid configuration, with $\Delta x = 0.25$, were $\sim 7\%$ and $\sim 4\%$ larger than their respective AMR configurations (Figure \ref{fig:FGamp1convergence}). For $A=8$, the AMR convergence test in Figure \ref{fig:AMRamp8convergence} shows that $E_\mathrm{massive}$ and $E_\mathrm{massless}$ converge at approximately third and fourth order respectively. The AMR simulations reach a maximum refinement level of $l_\mathrm{max}=4$ with $\Delta x = 0.0625$, and $E_\mathrm{massive}$ and $E_\mathrm{massless}$ measured by the finest fixed grid configuration, with $\Delta x = 0.25$, were $\sim 30\%$ and $\sim 6\%$ larger than their respective AMR configurations (Figure \ref{fig:FGamp1convergence}). The AMR convergence test for $A=1$ in Figure \ref{fig:AMRamp1convergence} shows that $E_\mathrm{massive}$ and $E_\mathrm{massless}$ converge at approximately sixth and fourth order respectively. The AMR simulations reach a maximum refinement level of $l_\mathrm{max}=3$ with $\Delta x = 0.125$, and $E_\mathrm{massive}$ and $E_\mathrm{massless}$ measured by the finest fixed grid configuration, with $\Delta x = 0.25$, were $\sim 86\%$ and $\sim 48\%$ larger than their respective AMR configurations (Figure \ref{fig:FGamp1convergence}). As the signals are already small for this low amplitude configuration, this error does not significantly affect our results (see Figure \ref{fig:AGGRadiation}). 




\begin{figure*}
    \centering
    \includegraphics[width=0.48\textwidth]{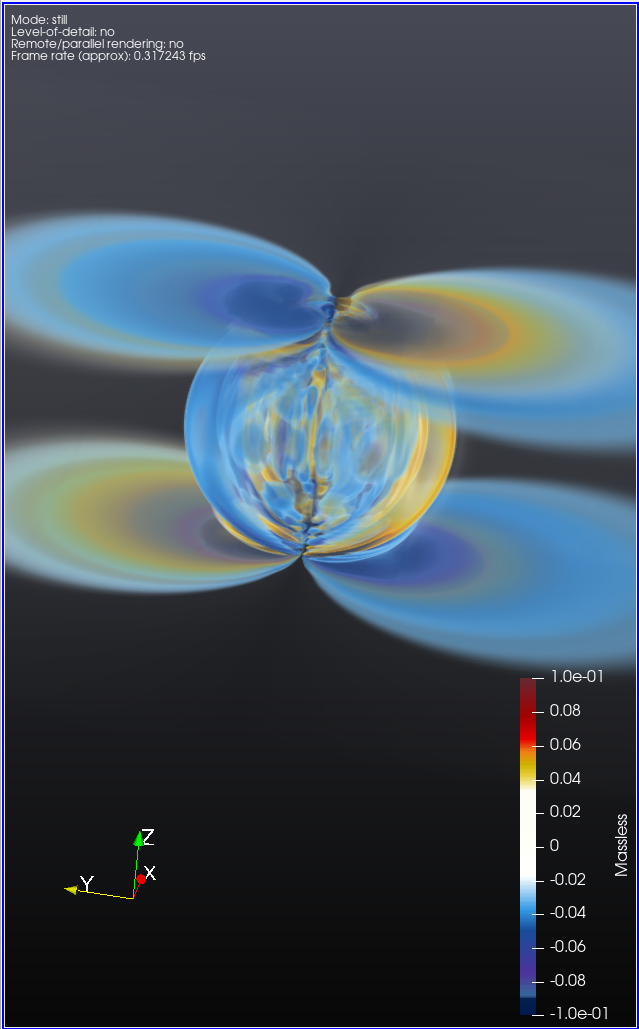}
    \includegraphics[width=0.48\textwidth]{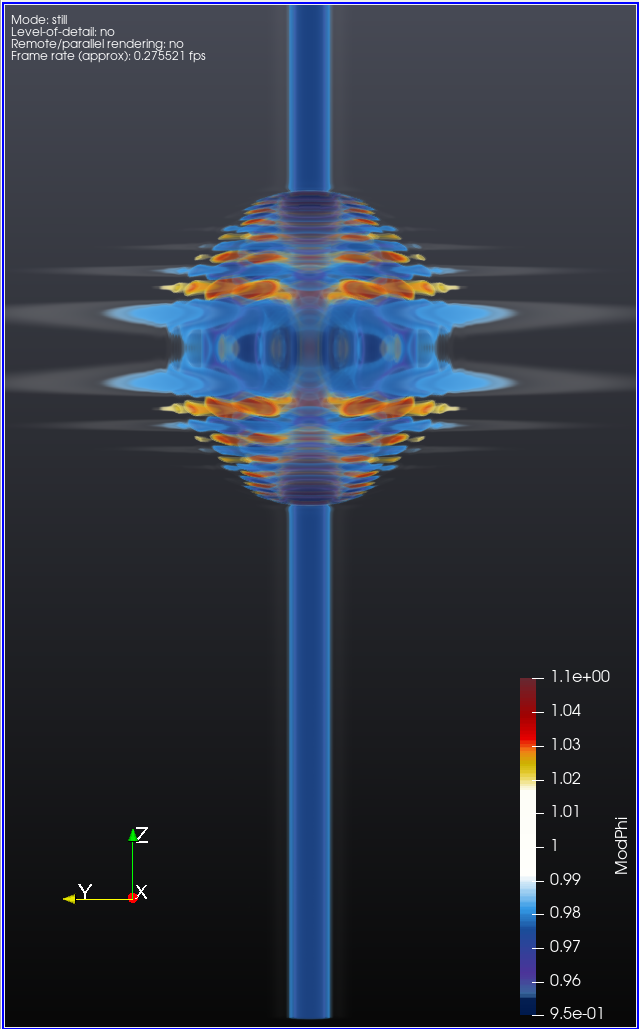}
    \caption{Volume rendering in 3D space $(x,y,z)$ of massless ${\Pi}_\vartheta$ radiation (left) and $|\phi|$ from a Gaussian/Gaussian travelling wave collision with $A=20$ and $\sigma_{\rm d}=2$, at $t\approx100$. The isotropic massless signal is qualitatively similar to the isotropic GW signal predicted from kink-kink collisions in \cite{Binetruy2009}.}
    \label{fig:screenshot}
\end{figure*}


\section{Massive and Axion Radiation}\label{massiveaxionradiation}

In this section, we plot results from over 100 simulations of colliding travelling wave configurations on an axion string. We perform parameter scans over initial amplitudes from $0.4 \leq A \leq 35$ with fixed standard deviations $\sigma_\mathrm{d}=2$ and $\sigma_\mathrm{d}=6$, and a parameter scan from $1 \leq \sigma_\mathrm{d} \leq 6$, with fixed amplitude $A=5$.  We consider two configurations of travelling waves: the collision of two Gaussians (G/G) (displacements with the same sign), and the collision of a Gaussian and an anti-Gaussian (aG/G) (opposite sign). For both, we consider initial amplitudes of equal magnitude for the travelling waves.

In light of the discussion of corresponding Nambu-Goto solutions in Section \ref{NambuGotostrings}, we define four regimes for the radiative channels that can be investigated with this parameter range (not necessarily mutually exclusive): 

\begin{enumerate}[label=\Roman*]


\item \textit{Quasi-linear Nambu-Goto regime, $R \gtrsim \sigma_{\rm d} \gg \cal{O}(\delta)$}, $A \leq A_{\mathrm{max}}$. These are travelling wave configurations with a large radius of curvature compared to the overall width of the travelling wave, i.e. a low amplitude $A$ relative to $\sigma_{\rm d}$, with $\sigma_{\rm d}$ larger than the string width. These closely approximate Nambu-Goto string solutions with left- and right-moving modes that smoothly pass through each other and an amplitude $A \le A_{\rm max} \approx \sigma_{\rm d}$ that is lower than the threshold calculated in Section \ref{NambuGotostrings}. These should reproduce Nambu-Goto results with ubiquitous massless radiation and highly suppressed massive emission. 

\item \textit{Nonlinear relativistic regime, $R < \sigma_{\rm d}$}, $A > A_{\mathrm{max}}$. These are super-relativistic travelling wave configurations with a small radius of curvature compared to the overall width of the source, i.e. a high amplitude $A$ relative to $\sigma_{\rm d}$ that is higher than the threshold calculated in Section \ref{NambuGotostrings}. These large amplitude collisions cause major energy loss through string annihilation and radiation. For our categorisation, this regime is agnostic to the ratio of $A$ or $\sigma_{\rm d}$ to the string width $\delta$.

\item \textit{Extreme topological regime, $A \gg A_{\rm max}$}.   These are hyper-relativistic travelling waves (more like `shock waves') where the large amplitude collisions are sufficiently energetic to create new loops which reconnect with the original string configurations.  Energy losses are reduced, but still very substantial. As with regime II, this regime is agnostic to the ratio of $A$ or $\sigma_{\rm d}$ to the string width $\delta$.

\item \textit{Above mass threshold, $\sigma_{\rm d} \approx \cal{O}(\delta)$}. These are narrow sources where the overall width of the travelling wave is comparable to the string width. Short wavelength excitations along the string $\sigma_{\rm d} \approx \delta$ mean that modest perturbations can excite massive modes $m_H \approx \delta^{-1}$.  For example,  $\sigma_{\rm d} =2$ is in this regime when  we choose parameters where $\delta =1$. This regime can coincide with regime II or III.

\end{enumerate}

Figure \ref{fig:screenshot} shows a 3D visualisation of the massless radiation diagnostic $\Pi_\vartheta$ \eqref{masslessdiagnostic} (left) and $|\phi|$ (right) from a G/G configuration in regimes III and IV, with $A=20$ and $\sigma_{\rm d} = 2$, after the travelling waves have collided. We observe the lobes of the massless self-field signal corresponding to the travelling waves moving away from the centre, along with an isotropic burst radiating from the point of collision, from which there are also subsequent internal mode oscillations of the string. Importantly, this massless spherical burst is qualitatively similar to the isotropic GW signal predicted from a kink-kink collision in the Nambu-Goto model \cite{Binetruy2009}. For $|\phi|$, we observe oscillations of the field along the length of the string with a clear ripple-like effect coming from a shock-like wave. Analysis of this and similar configurations forms the basis of this paper.

\begin{figure}
    \centering
    \includegraphics[width=0.5\textwidth]{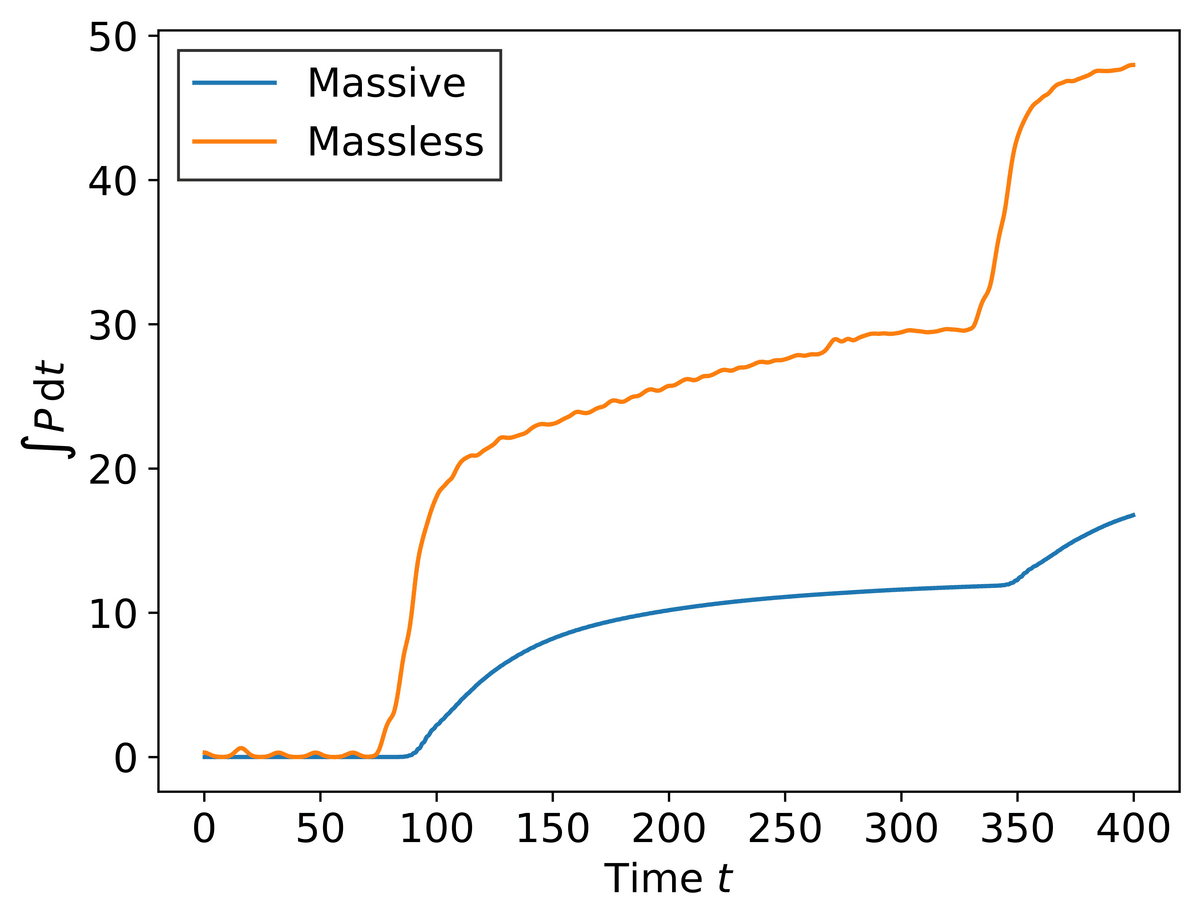}
    \caption{Time integral of the massive radiation signal $P_\mathrm{massive}$ \eqref{Pmassive} and massless radiation signal $P_\mathrm{massless}$ \eqref{Pmassless}, emitted from a Gaussian/Gaussian travelling wave collision with amplitude $A=5$ and $\sigma_{\rm d} = 3.5$.}
    \label{fig:example_signal}
\end{figure}

\begin{figure}[!b]
    \centering
    \includegraphics[width=0.5\textwidth]{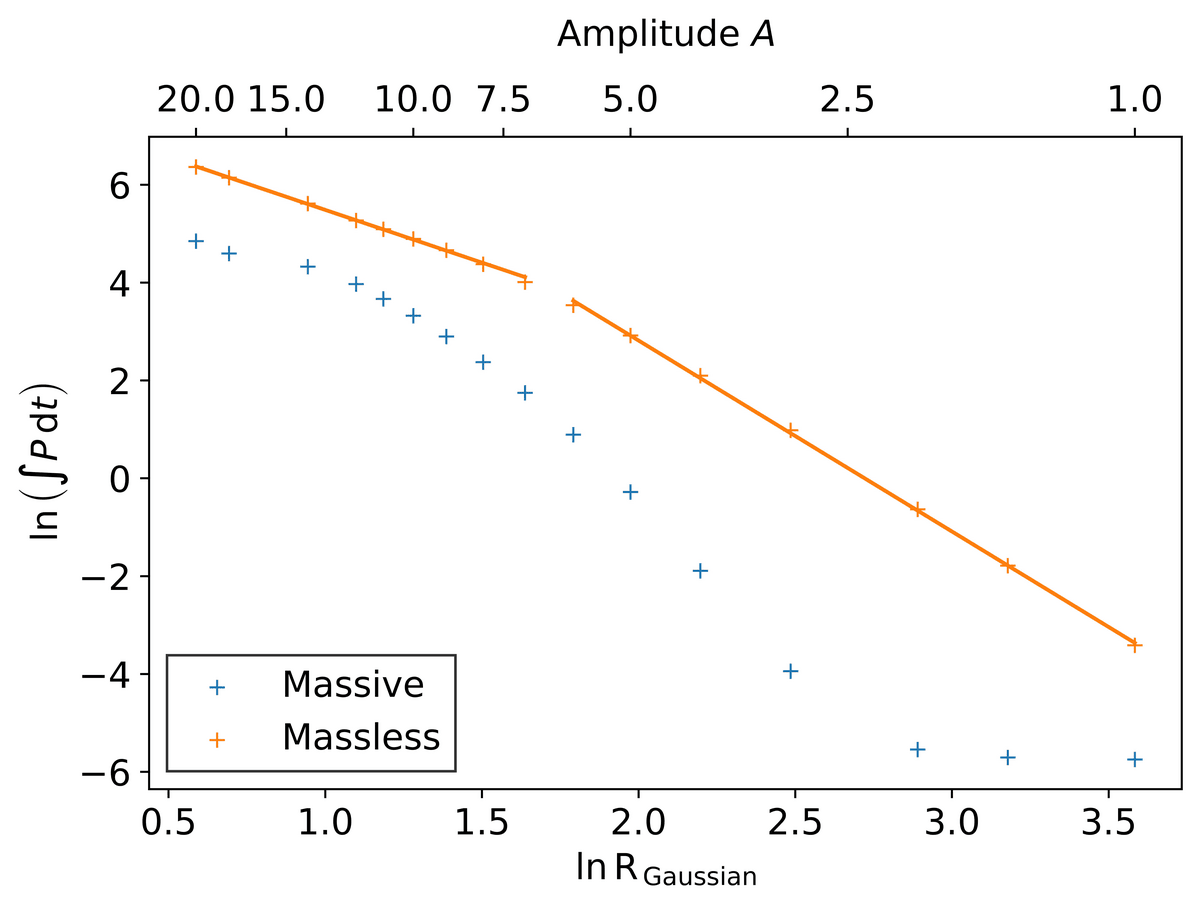}
    \includegraphics[width=0.5\textwidth]{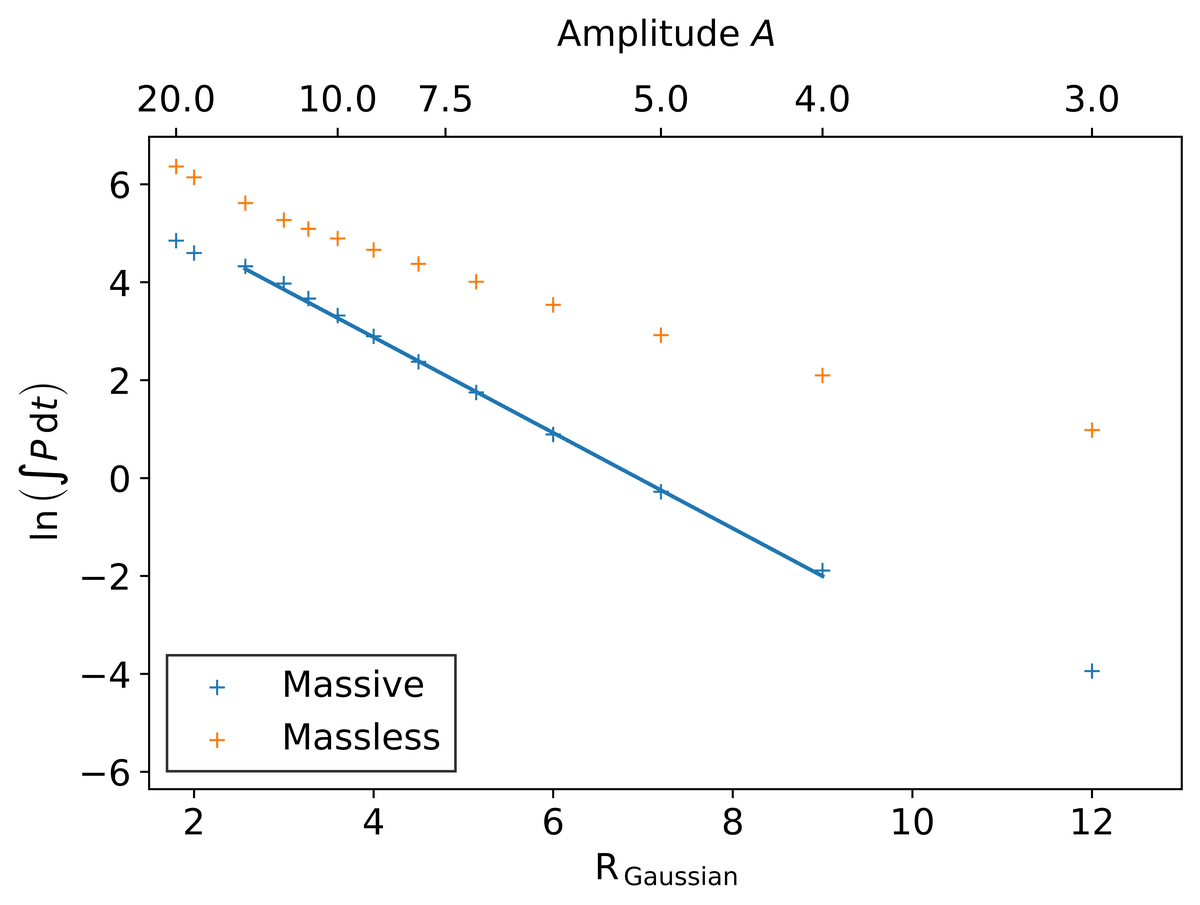}
   
    \caption{Massive (blue) and massless (yellow) radiation emitted from Gaussian/Gaussian collisions with amplitude $1 \leq A \leq 20$ and $\sigma_{\rm d} = 6$. The plots show the time integral of the massive and massless components of the Poynting vector, $P_\mathrm{massive}$ \eqref{Pmassive} and $P_\mathrm{massless}$ \eqref{Pmassless}, on the diagnostic cylinder at $R=64$ integrated from $t=0$ to $t=200$.}
    \label{fig:GGRadiationStdev6}
\end{figure}

\subsection{Absolute and Relative Magnitude}

We measure the time integral of the massive and massless radiation signals, $P_\mathrm{massive}$ \eqref{Pmassive} and $P_\mathrm{massless}$ \eqref{Pmassless}, for different burst configurations created by the collision of two travelling waves. An example of a signal from a G/G collision is given in Figure \ref{fig:example_signal}, which shows a burst of massless radiation reaching the extraction cylinder at $R=64$ at $t\approx 80$, with the massive signal arriving slightly afterwards. A second burst arrives after $t\approx 300$ which comes from the collision of the travelling waves with the periodic $z$ boundary. We analyse only the first burst signal in subsequent plots, choosing to integrate from $0 \leq t \leq 200$.

We concentrate first on a parameter scan that traverses the quasi-linear and highly nonlinear regimes I-III, but which are not in regime IV. Figure \ref{fig:GGRadiationStdev6} shows the magnitude of the massive and massless burst signals for G/G configurations with $\sigma_{\rm d} = 6 \gg \delta$ over a range of amplitudes $1 \leq A \leq 20$. We plot the dependence of these signals on the radius of curvature of the string at the centre of the initial travelling wave, $R_\mathrm{\,Gaussian}$. We use $R_\mathrm{\,Gaussian}$ rather than the radius of curvature at the point of collision because, for the aG/G configurations investigated later, the travelling waves will completely destructively interfere at this point, resulting in an instantaneously straight string with infinite radius of curvature that we cannot use as a model parameter. The radius of curvature along the string is defined by
\begin{equation}
    R = \frac{[1+X'(z)^2]^{3/2}}{X''(z)}\,,
\end{equation}
where $X(z)$ is given here by \eqref{StringCore} and denotes the position of the string core. Although this is not necessarily a Lorentz-invariant quantity, its definition here is unambiguous because we always work in the centre of mass frame in which the travelling waves collide. For a Gaussian, the radius of curvature is minimised at its peak, with a value of 
\begin{equation}
    R_\mathrm{\,Gaussian} = \frac{\sigma_{\rm d}^2}{A}\,.
\end{equation}
For fixed $\sigma_{\rm d}$ as in Figure \ref{fig:GGRadiationStdev6}, $R_\mathrm{\,Gaussian}$ is proportional to the inverse of the amplitude $A$. In the following discussion, we will also refer frequently to the `curvature' of the string defined as $1/R_\mathrm{\,Gaussian}$, as it is usually more intuitive to discuss an increase in curvature rather than a decrease in radius of curvature, and vice versa. We recall that the ratio of the radius of curvature $R$ to the string width $\delta$ determines the tension of the string via \eqref{tension}.

\begin{table}
    \centering
    \caption{Power-law coefficients $\gamma$ in different curvature regimes as defined in Section \ref{massiveaxionradiation} for Gaussian/Gaussian (G/G) and anti-Gaussian/Gaussian (aG/G) travelling wave configuration, defined by $E_{\mathrm{channel}} \propto (R_\mathrm{\,Gaussian})^{-\gamma}$. Values are given for the massive and massless decay channels separately, and the ratio between the two. Error bars are the statistical error from the linear regression.}
    \begin{ruledtabular}    
    \begin{tabular}{ccccc}\label{tab:results}
    Conf. & $\sigma_{\rm d}$ & Radiation & \multicolumn{2}{c}{Regime} \\
    \hline \hline
    & & & \multicolumn{2}{c}{$\sigma_{\rm d} \gg \delta$} \\
    \cline{4-5}
    & & & $A \gtrsim 6$ (II-III) & $A \lesssim 6$ (I) \\
    \cline{4-5}
    G/G& 6 &Massless & 2.15 $\pm$ 0.02 & 3.90 $\pm$ 0.04 \\       
    \hline \hline
    & & & \multicolumn{2}{c}{$\sigma_{\rm d} \lesssim 2 \delta$ (IV)}  \\
    \cline{4-5}
    & & & $A \gtrsim 6$ & $A \lesssim 6$ \\
    \cline{4-5}
    G/G& 2 &Massive & 1.61 $\pm$ 0.04 & 4.13 $\pm$ 0.03 \\
     & & Massless &  1.88 $\pm$ 0.05 & 3.79 $\pm$ 0.01\\
    aG/G & 2 & Massive &  0.44 $\pm$  0.12 & 3.55 $\pm$ 0.04  \\
     & & Massless  & 1.36 $\pm$ 0.03& 3.67 $\pm$ 0.03 \\
    \hline
     & & & \multicolumn{2}{c}{$A \gtrsim 2$} \\
    \cline{4-5}
    G/G & 2 & Massive/Massless  &  \multicolumn{2}{c}{1.00 $\pm$ 0.06}  \\
    aG/G & 2 & Massive/Massless  &  \multicolumn{2}{c}{0.87 $\pm$ 0.03} \\
    \end{tabular}
    \end{ruledtabular}
\end{table}

Figure \ref{fig:GGRadiationStdev6} shows that the magnitude of both the massive and massless burst signals increases as the amplitude (and therefore curvature) of the string increases (i.e. minimum local string tension decreases). This is as we would expect; it has been shown in previous publications that particle radiation is emitted more strongly from high curvature regions \cite{Drew2023, Blanco-Pillado_2023}, and that a higher amplitude also radiates more massless radiation \cite{Drew2019, Battye1993}. The massless radiation from the G/G configurations falls into two power law regimes; one corresponding to the nonlinear regimes II-III with high amplitude, and the other for the quasi-linear regime I with low amplitude. For $A/\delta \lesssim 6$, the radiation obeys $E_{\mathrm{massless}} \appropto A^4$, where we recall that $\eta$ and $\delta$ have been set to 1. The power law is quoted to one significant figure; the more precise gradient and statistical errors are given in Table \ref{tab:results}. 
For $A/\delta \gtrsim 6$, $E_{\mathrm{massless}} \appropto A^2$ as presented in Table \ref{tab:results}. 

\begin{figure}
    \centering
    \includegraphics[width=0.5\textwidth]{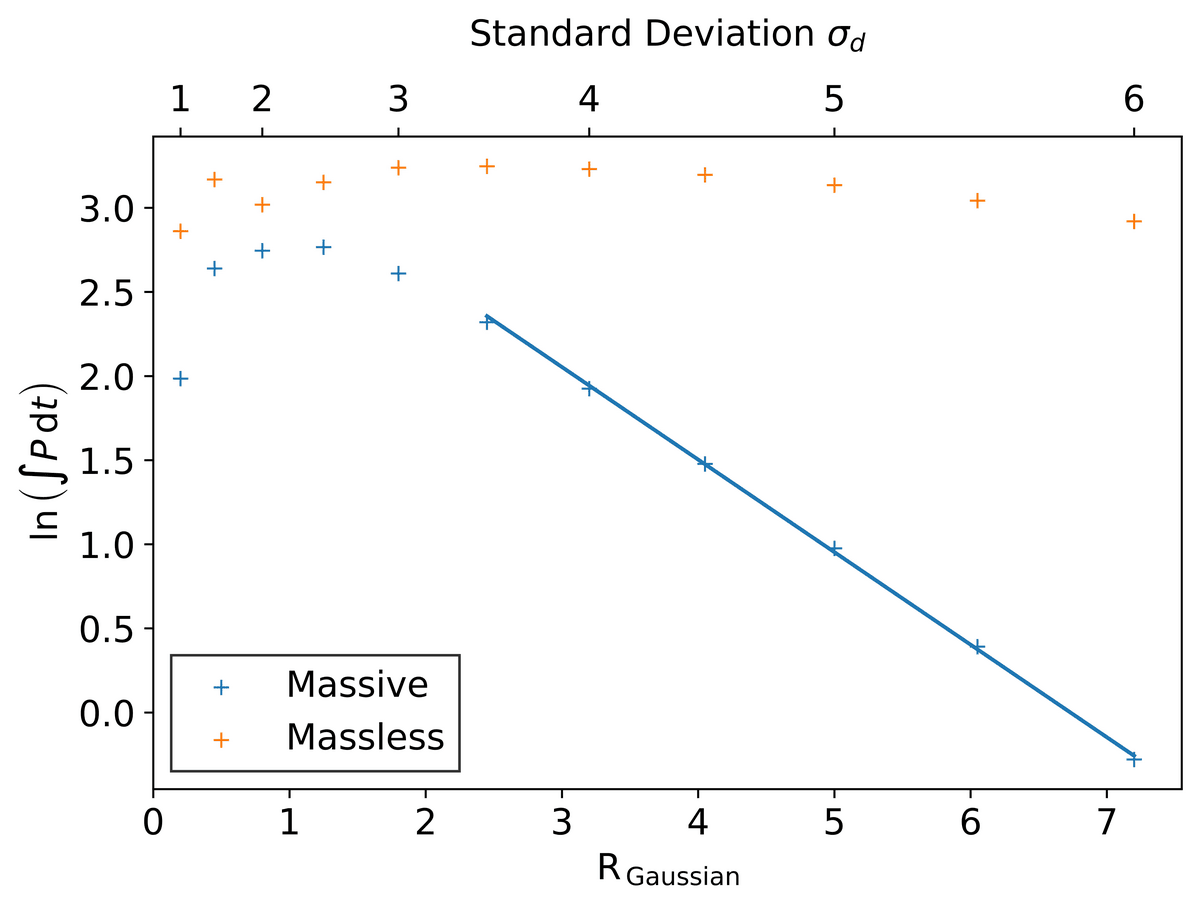}
    \caption{Massive (blue) and massless (yellow) radiation emitted from Gaussian/Gaussian collisions with amplitude $1 \leq \sigma_{\rm d} \leq 6$ and $A = 5$. The plots show the time integral of the massive and massless components of the Poynting vector, $P_\mathrm{massive}$ \eqref{Pmassive} and $P_\mathrm{massless}$ \eqref{Pmassless}, on the diagnostic cylinder at $R=64$ integrated from $t=0$ to $t=200$.}
    \label{fig:GGRadiationAmp5}
\end{figure}

Figure \ref{fig:GGRadiationAmp5} shows the results from a G/G parameter scan over $1 \leq \sigma_{\rm d} \leq 6$ with $A = 5$, incorporating all regimes. In this case, $E_{\mathrm{massless}}$ is only weakly dependent on the extent of the source; there is a slight suggestion that increasing $\sigma_{\rm d}$ decreases the overall magnitude of the radiation, but the effect is small. We therefore conclude that the massless radiation depends to leading order on $A/\delta$ alone, rather than the string curvature which includes $\sigma_\mathrm{d}$.

In contrast, Figures \ref{fig:GGRadiationStdev6} and \ref{fig:GGRadiationAmp5} show that the massive radiation is exponentially suppressed in the linear or quasi-linear regime I approximately as $E_{\mathrm{massive}} \appropto e^{-\zeta R_\mathrm{\,Gaussian}}$ where $\zeta$ is a dimensionful exponent with $[\zeta] = [E]$. Although both parameter scans exhibit exponential decay, the exponents have different values; for Figure \ref{fig:GGRadiationStdev6}, we obtain $\zeta = 0.98 \pm 0.02$ and Figure \ref{fig:GGRadiationAmp5} obtains $\zeta = 0.55 \pm 0.01$.  These reflect two different rescalings of the string configuration at the intersection of regimes I and IV, i.e. one changing amplitude $A$ at fixed $\sigma_{\rm d}$, and vice versa.  While the results are qualitatively consistent, more detailed analysis may be required to guide and improve our oversimplified modelling. It is possible, for example, that something closer to $E_{\mathrm{massive}} \appropto e^{-\,\zeta \sigma_\mathrm{d}^\alpha A^{-\beta}}$ with separate exponents $\alpha$ and $\beta$ for $\sigma_\mathrm{d}$ and $A$ respectively, would be more appropriate, potentially with the inclusion of additional constants. This will require more in-depth analysis and further simulations to determine, and is deferred to future work. 

In Figure \ref{fig:GGRadiationStdev6}, the massive radiation levels off at small amplitude $A \lesssim 3$ to become an approximately constant and very low value. We believe this is a numerical artifact related to the massive radiation at this amplitude not being  properly resolved by the numerical AMR grid (see discussion in Section \ref{convergence}). These outlier points are therefore not included in the bottom plot of Figure \ref{fig:GGRadiationStdev6}. Figure \ref{fig:GGRadiationAmp5} also includes a region $\sigma_\mathrm{d} \lesssim 2$ where $E_{\mathrm{massive}}$ does not exhibit exponential decay. In this highly nonlinear region, the initial Gaussian travelling waves are are so curved that the string is already significantly overlapping with itself prior to the collision (i.e. in regime IV). We therefore expect some of the energy of the configuration to already have dissipated before the travelling waves have a chance to collide, reducing the magnitude of the burst signal (see also the suppression that occurs in regime III).

\begin{figure}[!b]
    \centering
    \includegraphics[width=0.5\textwidth]{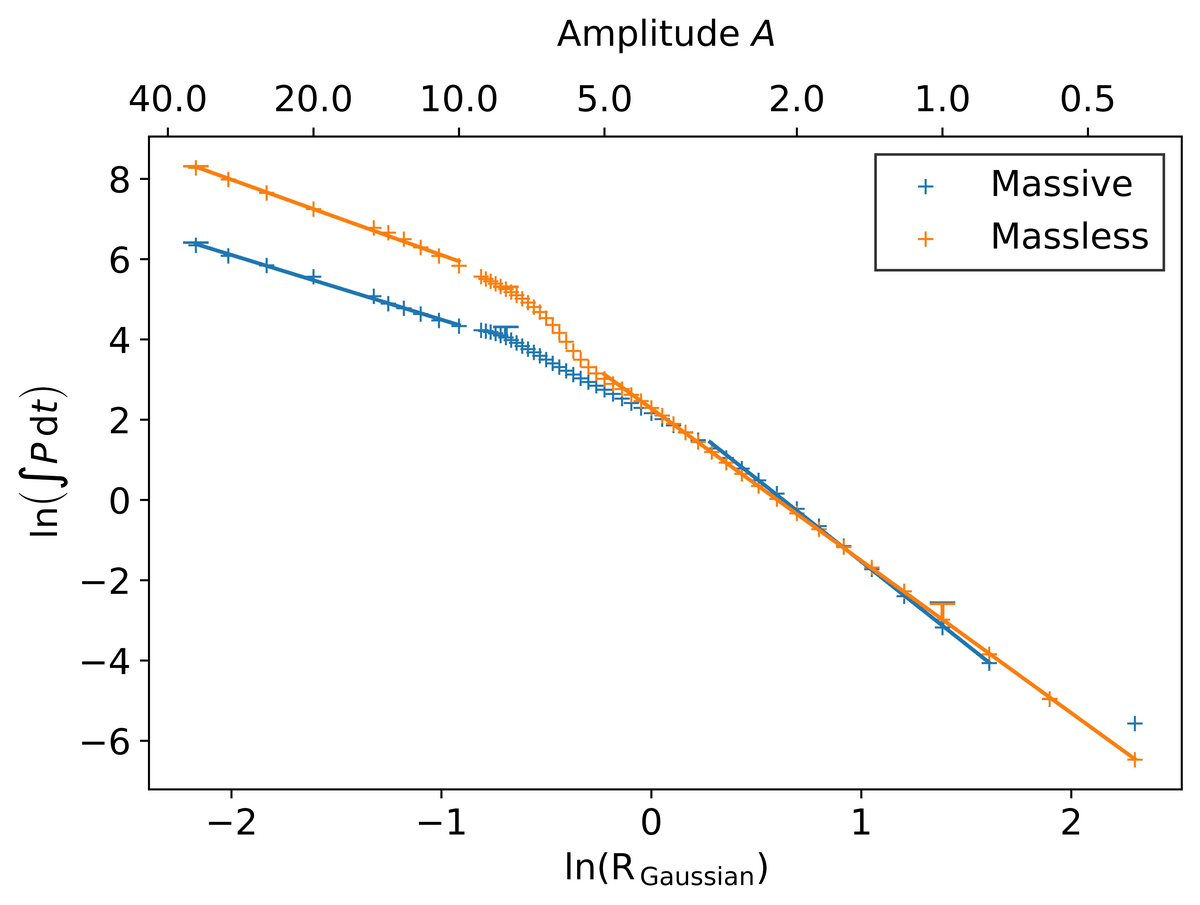}   
    \includegraphics[width=0.5\textwidth]{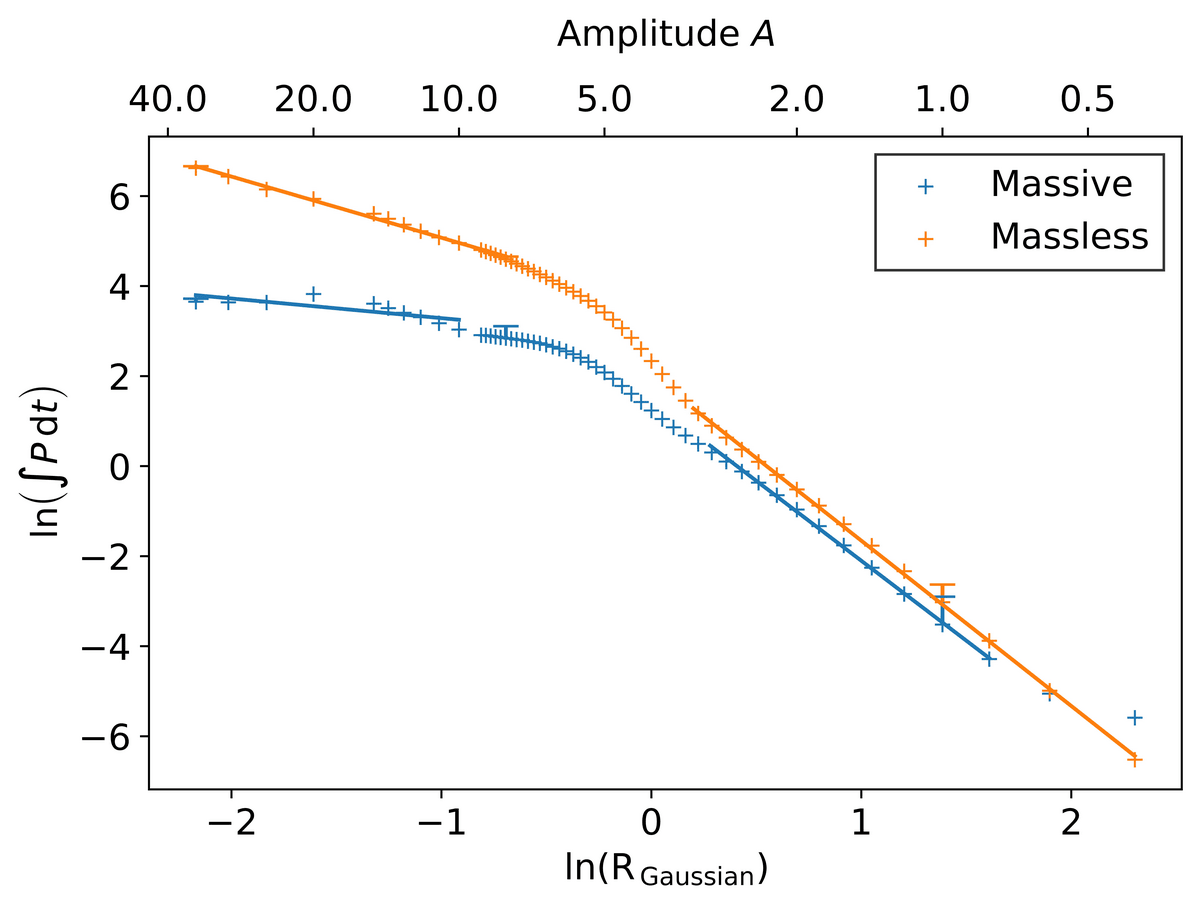}
    \caption{Massive (blue) and massless (yellow) radiation emitted from Gaussian/Gaussian collisions (top) and anti-Gaussian/Gaussian collisions (bottom) with amplitude $0.4 \leq A \leq 35$ and $\sigma_\mathrm{d} = 2$. The plots show the time integral of the massive and massless components of the Poynting vector, $P_\mathrm{massive}$ \eqref{Pmassive} and $P_\mathrm{massless}$ \eqref{Pmassless}, on the diagnostic cylinder at $R=64$ integrated from $t=0$ to $t=200$. Lines of corresponding colours indicate power law fits to the data for two regimes either side of the central `transition region'. The upper error bars are estimated for three points by comparing to fixed grid simulations.}
    \label{fig:AGGRadiation}
\end{figure}

We next concentrate on parameter scans over the nonlinear to highly nonlinear (extreme topological) regimes II and III. This figure also covers regime IV, because $\sigma_\mathrm{d} \lesssim 2\delta$, then includes regime II with $A/\delta \lesssim 6$ and III with $A/\delta \gtrsim 6$ respectively. Figure \ref{fig:AGGRadiation} plots results for G/G and aG/G travelling wave configurations, with upper error bars estimated using the fixed grid comparison runs in Section \ref{convergence}. Looking first at the massless radiation, similarly to Figure \ref{fig:GGRadiationStdev6}, there is a distinct change in the power law which occurs around the same value of $A \approx 6$, despite the different values of $R_{\,\mathrm{Gaussian}}$. In this case, we observe a more extended `transition region' between $4 \lesssim A \lesssim 9$. The massless radiation either side of the transition region obeys approximately equivalent power laws to the regions either side of $A \approx 6$ in Figure \ref{fig:GGRadiationStdev6}, $E_\mathrm{massless} \appropto A^4$ in the lower amplitude region and $E_\mathrm{massless} \appropto A^2$ in the higher amplitude region, although we note that this is closer to $E_\mathrm{massless} \appropto A$ for the high amplitude aG/G configuration. We also note that the values of the power laws do not match between the corresponding \textit{curvature} regimes of the two datasets. This consolidates our conclusion that the magnitude of the massless radiation depends to leading order only on $A/\delta$, with little explicit dependence on $R$.

In the highly nonlinear regime II with $A \lesssim 6$, the massive radiation is present at comparable magnitude to the massless radiation and is described by a similar power law $E_{\mathrm{massive}} \appropto A^4$, given more precisely in Table \ref{tab:results}. The massive radiation is slightly suppressed for the aG/G configuration in the regime $A \lesssim 6$ relative to the G/G configuration. This makes sense as, for the aG/G configuration at the point of collision, the string is instantaneously flat rather than highly curved as in the G/G case. Configurations with very low amplitudes $A \lesssim \delta$ emit a higher magnitude of massive radiation than massless, in contrast with the rest of the parameter space. However, this is likely to be due to the effects of imperfect initial conditions becoming dominant. 

In the extreme topological regime III with $A \gtrsim 6$, the massive radiation is relatively more suppressed, although it still follows a similar power law to the massless radiation. In fact, both the massive and massless channels are perhaps suppressed compared to if we were to naively extrapolate from regime II. As we will see, this is likely due to the formation of new loop structures from coherent radiation from these extremely high energy configurations, which has the effect of relatively damping the radiation signal as the energy instead forms new string. For the G/G configuration, the massive radiation power law relationship in the high amplitude region $E_\mathrm{massive} \appropto A^2$ is consistent with the corresponding amplitude regime in Figure \ref{fig:GGRadiationStdev6}, where at high $A$ the massive radiation tends towards the same power law as the massless radiation. However, the value of the power law does not agree in the same \textit{curvature} regime between the two scans. This lends weight to the previous suggestion that there is some additional dependence on $A$ or $\sigma_\mathrm{d}$ for the massive radiation aside from $R_{\,\mathrm{Gaussian}}$, although in this regime we are looking at a power law relationship, rather than an exponential suppression. Further interesting features of the highly nonlinear regime in Figure \ref{fig:AGGRadiation} are that both radiation channels are suppressed for the aG/G configurations with respect to the G/G configurations, and the power law for the G/G case is also steeper for both channels.

\begin{figure*}
\centering
    \includegraphics[width=\textwidth]{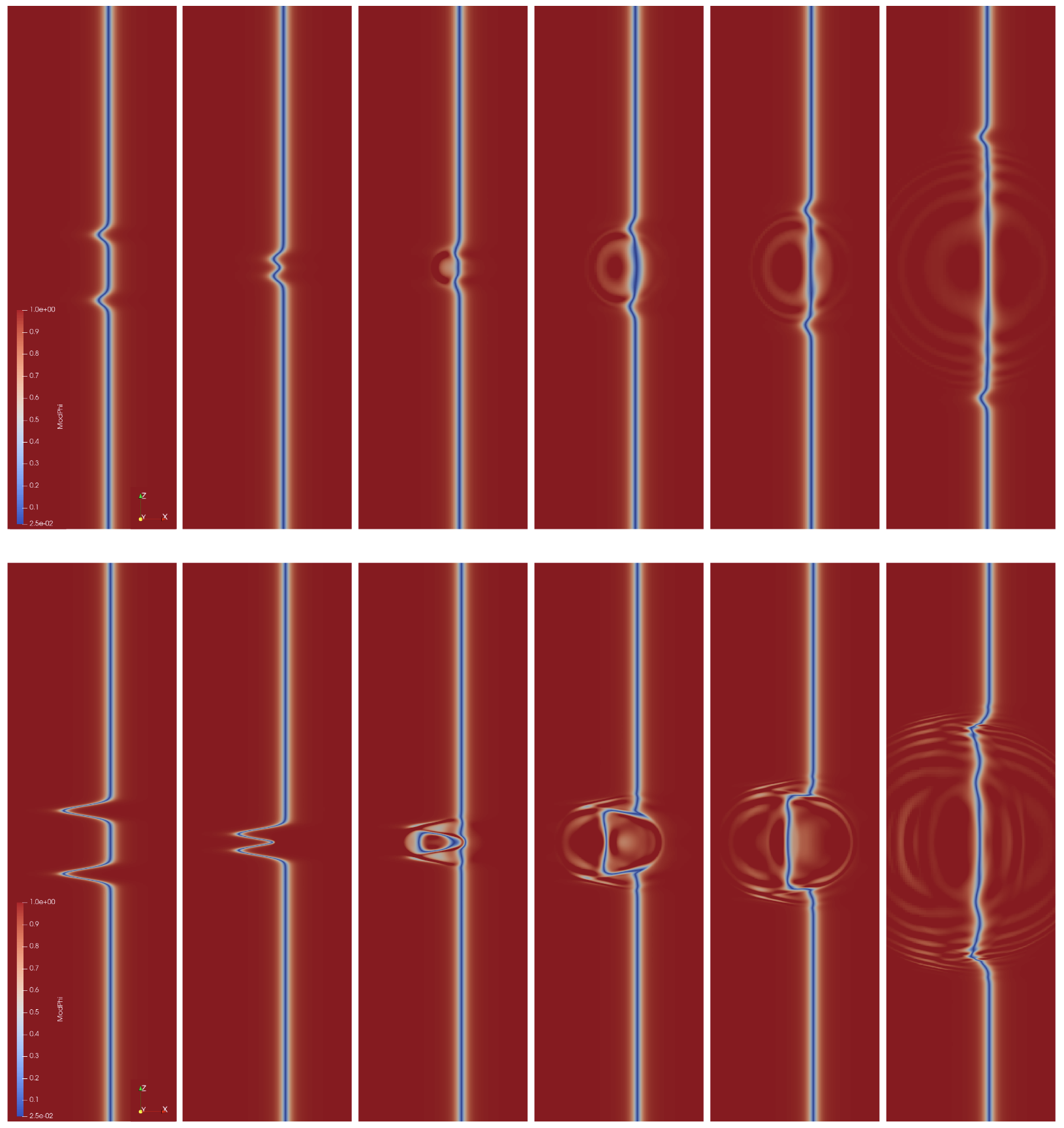}
    \caption{2D slices through two Gaussian/Gaussian travelling wave configurations, with snapshots over time running from left to right. The top configuration shows $|\phi|$ from $A=4$ and $\sigma_\mathrm{d}=2$, and the bottom configuration is $A=20$ and $\sigma_\mathrm{d}=2$}
    \label{nonlinear}
\end{figure*}


Figure \ref{nonlinear} shows an example evolution for regime II with $A \lesssim 6$ and regime III with $A \gtrsim 6$, both in regime IV with $\sigma_\mathrm{d} = 2$. We observe fundamentally different qualitative behaviour of the travelling waves between the two regimes. For the lower amplitude $A=4$, the two travelling waves collide, pass through each other and move apart, with internal mode oscillations generated at the point of collision continuing to radiate massive radiation. For the highly nonlinear $A=20$, we observe some very extreme phenomena; at the point of collision, the travelling waves have sufficient energy that they instantaneously create an additional loop of string, which then reconnects with the long string in such a way that the central portion of the string is displaced, which we could potentially interpret as a `memory' effect. The extreme topological regime of $A \gtrsim 6$ with $\sigma_\mathrm{d} = 2$ is therefore fundamentally different from the behaviour we observe in the other regimes, even compared to other nonlinear configurations in regime II.


\begin{figure}
    \centering
    \includegraphics[width=0.5\textwidth]{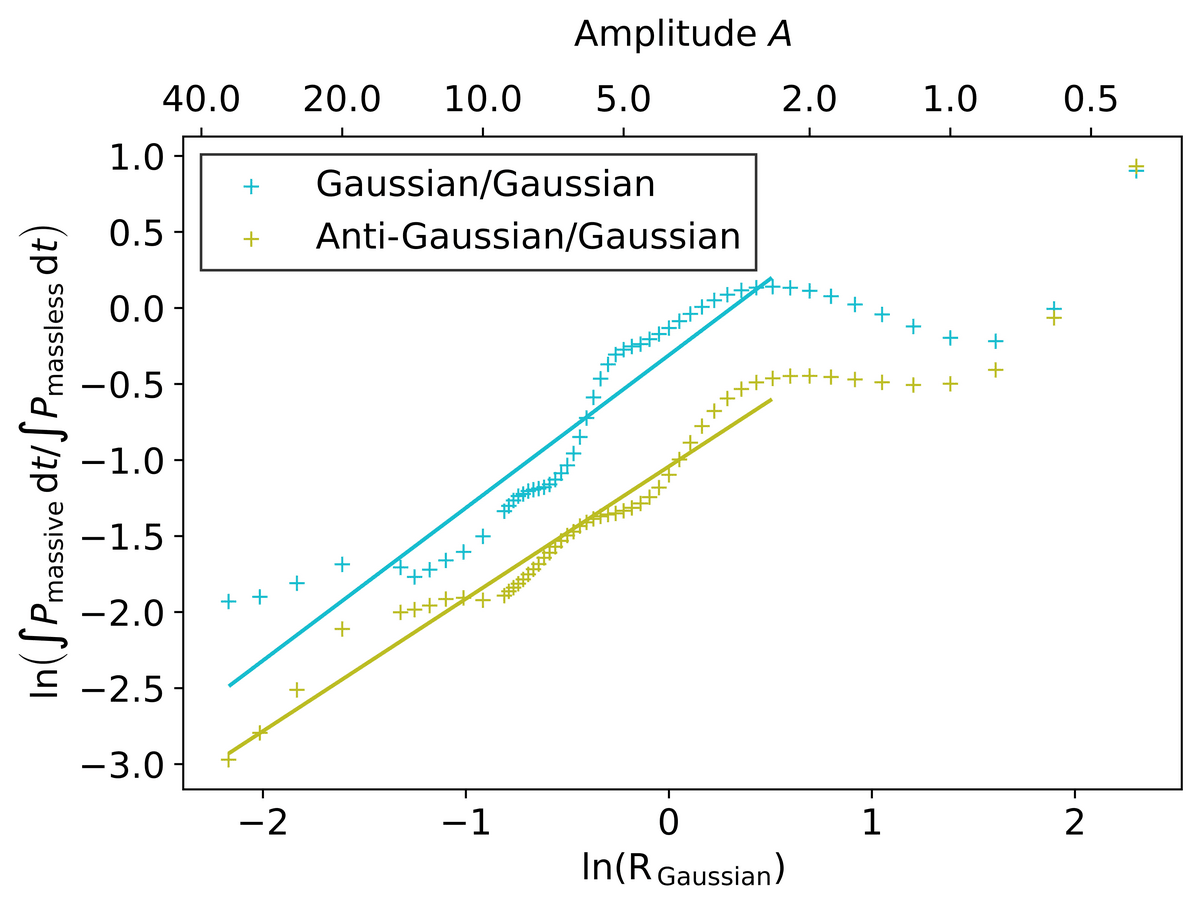}    
    \caption{Ratio of massive to massless radiation emitted from Gaussian/Gaussian and anti-Gaussian/Gaussian collisions with amplitude $0.4 \leq A \leq 35$ and $\sigma_\mathrm{d} = 2$. The plot shows the ratio of the time integrated massive to massless components of the Poynting vector, $P_\mathrm{massive}$ \eqref{Pmassive} and $P_\mathrm{massless}$ \eqref{Pmassless}, on the diagnostic cylinder at $R=64$ integrated from $t=0$ to $t=200$. Lines of corresponding colours indicate power law fits to the data for the high curvature regime.}
    \label{fig:AGGRatio}
\end{figure}

\begin{figure}
    \centering
    \includegraphics[width=0.5\textwidth]{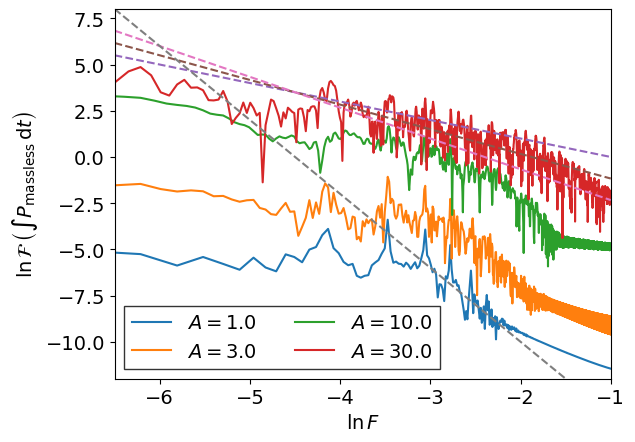}
    \caption{Spectrum of the massless radiation $P_{\mathrm{massless}}$ emitted from Gaussian/Gaussian travelling wave configurations with a range of $A$. We observe a decrease of the spectral index (more UV) with increasing amplitude. Dashed lines correspond to spectral indices of 1 (purple), 4/3 (brown), 5/3 (pink) and 4 (grey).}
    \label{fig:spectrum}
\end{figure}

Finally, Figure \ref{fig:AGGRatio} shows the dependence of the ratio of the massive to massless radiation for the same range of $R_{\mathrm{\,Gaussian}}$ as plotted in Figure \ref{fig:AGGRadiation}. We observe for $A \gtrsim 2$ that the ratio of radiation channels for the G/G configurations is consistent with a linear dependence on $R_{\mathrm{\,Gaussian}}$, i.e.
\begin{equation}
    \frac{E_\mathrm{massive}}{E_\mathrm{massless}} \propto R_{\mathrm{\,Gaussian}}\,.
\end{equation}
We also observe from Figure \ref{fig:AGGRatio} that for $1 \lesssim A \lesssim 2$, the massive to massless ratio flattens out to approximately a constant, or a slight decrease with $R_{\,\mathrm{Gaussian}}$ for the G/G configuration. The ratio then increases again for $A \lesssim 1 \approx \delta$, i.e. amplitudes smaller than the string width. Here, the overall magnitude of the radiation is so small that it is not clear that this region needs to be considered. However, if this region is physically relevant, it means that for very small, linear displacements of the string, the massive radiation in fact becomes comparable to or larger than the massless channel, whilst we are in regime IV.



\subsection{Axion Radiation Spectrum}

In Figure \ref{fig:spectrum}, we plot the massless radiation spectrum from G/G configurations with $A=1, 3, 10\; \mathrm{and} \;30$. We plot on a log-log scale to enable a power law fitting, along with some example power laws to guide the eye. For $A=1$, a power law fit of $\ln\mathcal{F}(E_\mathrm{massless}) \propto -\,4 \ln {f}$ approximates the high frequency part of the spectrum. For higher amplitudes such as $A=30$, the spectrum is approximated more accurately by $\ln\mathcal{F}(E_\mathrm{massless}) \propto -\ln {f}$. The power laws $\ln\mathcal{F}(E_\mathrm{massless}) \propto -\,\frac{4}{3} \ln {f}$ and $ -\,\frac{5}{3} \ln {f}$ also provide a plausible fit, where we choose these values bearing in mind the analytic analogy between massless axion radiation from global strings and gravitational radiation from Abelian-Higgs strings (see \nameref{AppendixC}) and the GW spectrum predictions for cusps and kinks \cite{Vilenkin:2000jqa, Damour2001, Damour2005}. In general, the spectral index, defined here to be $q$ where $\mathcal{F}\left(E_{\mathrm{massless}}\right) \propto F^{-q}$ and $F$ is frequency, decreases as the amplitude and curvature increases. This means that there is a higher proportion of radiation emitted in high frequency modes as the curvature increases. The implications of these spectral index measurements are discussed further in Section \ref{spectrumdiscussion}.

\section{Comparison with a Periodic Source}\label{Discussion}

In this section, we link the results obtained in Section \ref{massiveaxionradiation} to those obtained for sinusoidal string configurations in \cite{Drew2019, Drew2023}. In \cite{Drew2019, Drew2023}, sinusoidal configurations of axion strings were analysed using similar techniques to those used in this travelling wave investigation, allowing us to compare the two configurations to make more general statements about the dependence of the string radiation on the string parameters. 

We first note that it is not possible to meaningfully compare the \textit{magnitude} of the massive radiation using the current modelling, due primarily to the fact that the magnitude of the radiation also depends either on a currently undetermined prefactor, a more subtle dependence on $A$ and the source length $\sigma_\mathrm{d}$, or both. Even within the travelling wave investigation, if we compare, for example, the magnitude of the massive radiation from $\ln R_{\,\mathrm {Gaussian}} \approx 1$ in Figures \ref{fig:GGRadiationStdev6} and \ref{fig:AGGRadiation}, there is a difference of $\mathcal{O}(100\times)$, indicating an additional dependence on $A$ and/or $\sigma_\mathrm{d}$ that is not yet included in the model. This prefactor and further details of the parameter dependence will be important if we wish to make more general statements about the massive radiation from different configurations.

In contrast, comparing the magnitude of the massless radiation for the same two Figures \ref{fig:GGRadiationStdev6} and \ref{fig:AGGRadiation}, it does appear as though radiation of a comparable magnitude is emitted for configurations with the same amplitudes, which may facilitate a meaningful comparison to the sinusoidal case. We must bear in mind that the travelling wave configuration only emits one burst signal, when the Gaussians `collide' at $z=0$. In contrast, for a sinusoidal configuration, the string oscillates periodically, performing a full oscillation slightly less than once every time period $\Delta t =L$, where $L$ is the wavelength of the source. In this case, each oscillation has \textit{four} incidences of maximum amplitude; at $z=L/4$ and $3L/4$, twice each per oscillation. We also bear in mind that the amplitude of the sinusoidal oscillations decreases as the simulation progresses due to damping.

\begin{figure}
    \centering
    \includegraphics[width=0.5\textwidth]{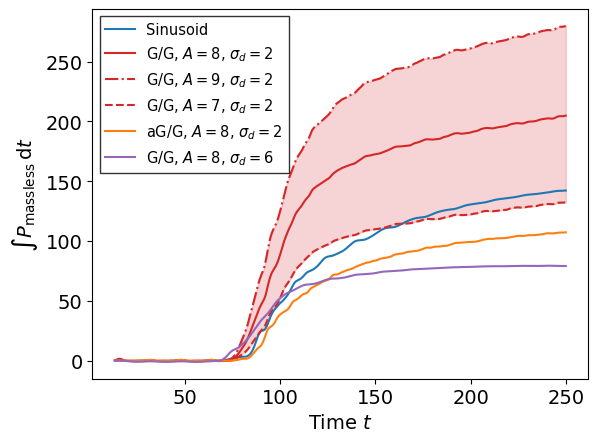}
    \caption{Cumulative integral of the massless radiation signal $P_\mathrm{massless}$ \eqref{Pmassless} emitted from a $\lambda=1$ string from \textit{i)} a sinusoidal configuration with $A=8$ and $L=32$ (blue), \textit{ii)} a Gaussian/Gaussian burst configuration with $A=8$ and $\sigma_\mathrm{d}=2$ (red), \textit{iii)} an anti-Gaussian/Gaussian burst configuration with $A=8$ and $\sigma_\mathrm{d}=2$ (orange) and \textit{iv)} a Gaussian/Gaussian burst configuration with $A=8$ and $\sigma_\mathrm{d}=6$ (purple). We also plot the signals from Gaussian/Gaussian burst configurations with $A=9$ (red dashed) and $A=7$ (red dash-dotted), both with $\sigma_\mathrm{d}=2$.}
    \label{fig:comparison}
\end{figure}

Figure \ref{fig:comparison} shows the cumulative integrated massless signal $E_\mathrm{massless}$ emitted by the two configurations. We plot the travelling wave signal for $A=8$ with $\sigma_\mathrm{d}=2$ for both the G/G and aG/G configurations and $A=8$ with $\sigma_\mathrm{d}=6$ for the G/G configuration, along with the sinusoidal signal for $A=8$ and $L=32$. We also plot the G/G travelling wave signals for $A=7$ and $A=9$ with $\sigma_\mathrm{d}=2$ to provide a comparison with nearby amplitudes. The $A=8$ sinusoidal signal is approximated well by all of the $A=8$ travelling wave configurations to within 50\% of its total magnitude. Of the $A=8$ configurations, the aG/G travelling wave signal fits closest, which makes sense if we consider its similarity with the sinusoidal string close to the point of collision. We also note that the $A=7$ with $\sigma_\mathrm{d}=2$ configuration approximates the signal better than $A=8$ with $\sigma_\mathrm{d}=2$. This also makes sense, if we recall that the sinusoidal configuration decreases in amplitude via damping over time. We conclude that the amplitude of the string provides a useful order of magnitude estimate of the magnitude of the massless radiation emitted, which is likely to be of use both for network and individual source modelling.

For the sinusoidal case in \cite{Drew2019, Drew2023}, the magnitude of the massive and massless radiation was determined for different widths of string, where the width was altered by changing the value of the parameter $\lambda$. As detailed in Section \ref{AxionStringTheory}, rescaling by $\sqrt{\lambda}$ is equivalent to scaling the radius of curvature $R$. This means that, by performing a parameter scan over $\lambda$, \cite{Drew2019, Drew2023} also effectively performed a scan over the maximum radius of curvature $R_{\,\mathrm{sin}}$ of the string. We can therefore compare the travelling wave parameter scan over radius of curvature with the sinusoidal $\lambda$ parameter scan, to determine whether or not the observations are consistent. The highest curvature configuration studied in \cite{Drew2023} with $A=6$ and $L = 16 \gg \delta$, lies either in regime II or III. A power law decay is obtained in \cite{Drew2023} for the massive radiation for this configuration, with a coefficient of $\gamma_{\mathrm{pow}} = 2.1\; (\pm 0.1)$. This is consistent with regime III in Figure \ref{fig:AGGRadiation}. For the sinusoidal configuration with $A = 8$ and $L=32$, a power law coefficient of $\gamma_{\mathrm{pow}} = 6.2\; (\pm 0.2)$ was obtained. This is not directly consistent with any power law obtained in the travelling wave study, providing further evidence that our massive radiation model requires refinement.

\section{Application to Gravitational Wave Modelling}\label{spectrumdiscussion}

In general, the modelling of gravitational waveforms in general relativity relies on a combination of analytic modelling of linear perturbations, and numerical simulation of the strong-field regime. This has been used successfully to determine the `chirp' signal expected for binary black hole mergers, where the `inspiral' phase is modelled using perturbation theory, and the `merger' is usually determined by numerical relativity simulations or using post-Newtonian approximations. In the case of cosmic strings, waveform modelling of burst signals has been carried out primarily using analytic models (for a review, see \cite{Vilenkin:2000jqa, Siemens2006}). These are useful in linear regimes but, in principle, they may fail to accurately capture the dynamics in the strong field regime, i.e. in regions of highly relativistic regions with high curvature and acceleration. 

The most widely studied burst signal configurations from cosmic string networks are the Nambu-Goto `cusp' and `kink' configurations for Abelian-Higgs strings. In the Nambu-Goto model, the gravitational waveform can be phenomenologically described by a simple power law shape
\begin{equation}
    h_q(\ell, z, f)=A_q(\ell, z) f^{-q} \Theta\left(f_h-f\right) \Theta\left(f-f_{\ell}\right)\,,
\end{equation}
where $q = 4/3$, $5/3$ and $2$ for cusps, kinks and kink-kink collisions respectively, $f$ is frequency, $A_q(\ell, z)$ is the amplitude, $\ell$ is the loop size and $z$, the cosmological redshift. The amplitude is given by
\begin{equation}
    A_q(\ell, z)=g_1 \frac{G \mu \ell^{2-q}}{(1+z)^{q-1} r(z)}\,,
\end{equation}
where $r(z)$ is the proper distance to the source and $g_1$ is an `uncertainty' factor, which is usually set to unity. This waveform has been used to search for cusp and kink burst signals in the first Advanced LIGO observing run \cite{Abbott2018} and subsequently in the third Advanced LIGO–Virgo observing run \cite{Abbott2021}.

Neglecting (for now) the dependence on redshift and the cutoff frequencies, which are determined by the beaming angle and the size of the burst-emitting configuration, the parameters that determine the gravitational waveform in this model are the string tension, $G\mu$, and the loop size $\ell$. The dependence on the string tension is intuitive; for a higher $G\mu$, strings have a higher mass per unit length, and therefore can be expected to emit a stronger signal than a lighter strings when accelerating near the speed of light. The dependence on $\ell$ is a more direct consequence of the Nambu-Goto modelling, which assumes that the length of a string loop determines the actual discrete frequencies making up its waveform via the loop harmonics. It is for this reason that the loop distribution function has been so important in modelling cosmic string signals, setting the frequency spectrum, and so why different loop models lead to very different GW signal predictions.

It is also well-understood that this Nambu-Goto model has its limitations. First, we know that cosmic strings are not infinitely thin, despite this being a good approximation on cosmological scales. This is not so important for low curvature configurations, but in regions where the string curvature is high and comparable to the string width, Nambu-Goto modelling will become inaccurate. It has been shown that radiation backreaction in field theory simulations causes strings to diverge from Nambu-Goto behaviour in high curvature regions \cite{Blanco-Pillado_2023}. As we have seen in the present simulations, we expect such high energy configurations of physical cosmic strings to emit massive particle radiation, a nonlinear effect which is not captured in the Nambu-Goto model. It has been the subject of some discussion whether this particle radiation will significantly affect the overall network evolution on cosmological scales. However, particle radiation can be important microphysically for cusp or cusp-like configurations (see e.g. \cite{Auclair_2021}), as these are exactly the high curvature, nonlinear regions for which we expect to see massive radiation become a relevant decay channel. Massive particle emission could therefore have a backreaction effect on the magnitude of the GW signal. Some important questions become: to what extent does massive particle radiation modify the GW amplitude, and on what frequencies might we expect this to affect, or even cut off, GW burst signals from `cusp-like' configurations?

For the physical setup presented in this paper, we consider massless and massive particle radiation from global strings. Although we do not directly consider gravity, the behaviour of massless radiation from global strings has been shown to be approximately analagous to gravitational radiation from local Abelian-Higgs strings (see \nameref{AppendixC}). We therefore postulate that observations made in this study about the balance between the two channels may also apply to GW signals from local strings. For the high curvature (regime II) configurations plotted in Figure \ref{fig:AGGRadiation}, we observed collision and annihilation cases in which up to $50\%$ of the total signal was emitted in massive particle radiation. If such configurations arise generically in realistic networks, and the analogy between axions and GWs carries over, then we might expect particle radiation from such burst signals in this regime which would also reduce the relative amplitude of predicted GW signals. The details of the knock-on effects for GW observational constraints on cosmic strings are left to future work.

\section{Conclusion and Future Work}\label{Conclusion}

We have investigated massive and massless (axion) radiation emitted from axion string networks and cusp-like burst configurations using adaptive mesh refinement simulations. We have observed from network simulations that massive radiation is emitted most strongly from regions of high string curvature, in contrast to axion radiation, which is more diffuse. This has motivated an investigation into the dependence of both channels on the maximum string radius of curvature $R_{\,\mathrm{Gaussian}}$, which also determines the local string tension.

We have presented the results of parameter scans over $R_{\,\mathrm{Gaussian}} = \sigma_\mathrm{d}^2/A$ for colliding Gaussian/Gaussian (G/G) and anti-Gaussian/Gaussian (aG/G) travelling wave configurations, obtained by implementing initial conditions from \cite{Vachaspati1990}. We have performed scans over different non-mutually-exclusive parameter regimes, defined to be: I. Quasi-linear Nambu-Goto regime where $R \gtrsim \sigma_{\mathrm{d}} \gg \mathcal{O}(\delta)$ and $A \leq A_{\rm{max}}$, where $A_{\rm{max}} \approx \sigma_\mathrm{d}$ is the maximum amplitude permitted in the Nambu-Goto model \eqref{NGconstraint},
II. Nonlinear relativistic regime where $R < \sigma_\mathrm{d}$ and $A > A_\mathrm{max}$, 
III. Extreme topological regime where $A \gg A_{\rm{max}}$ and IV. Above mass threshold where $\sigma_\mathrm{d} \approx \delta$ and massive modes $m_H \approx \delta^{-1}$ can be excited.

Scans over $1 \leq A \leq 20$ with $\sigma_\mathrm{d}=6$ and $1 \leq \sigma_\mathrm{d} \leq 6$ with $A=5$ showed that the energy emitted in massless radiation is consistent with a power law relationship $E_{\mathrm{massless}} \propto A^{4}$ in regime I/II with $A \lesssim 6$, and $E_{\mathrm{massless}} \propto A^{2}$ in regime II/III $A \gtrsim 6$ (with the latter scan also regime IV at low $\sigma_{\mathrm{d}})$. The scan over $\sigma_\mathrm{d}$ at fixed $A$ showed no significant dependence of massless radiation on the spatial extent of the source $\sigma_\mathrm{d}$. Massive radiation was shown to be exponentially suppressed with $E_{\mathrm{massive}} \propto e^{-\zeta R_{\mathrm{\,Gaussian}}}$ in regime I/II, although it was noted that the exponent $\zeta$ was not consistent between the scans. This indicates that our very simple model for massive radiation requires further devlopment to incorporate such extreme nonlinear effects.

In a scan over the nonlinear regimes II and III for $0.4 \leq A \leq 35$ with $\sigma_\mathrm{d}=2$ (i.e.\ also in regime IV), we found that the magnitude of both decay channels is consistent with two power laws linked by a more extended `transition region' around $4 \lesssim A \lesssim 9$. For both channels, the power law in regime III for the G/G configuration is consistent with an inverse square law $E \propto R_{\,\mathrm{Gaussian}}^{-2}$ to one significant figure, and regime II is consistent with $E \propto R_{\,\mathrm{Gaussian}}^{-4}$, where the power laws have slightly lower exponents for the aG/G configurations. We note that, for the massless radiation, the power laws are approximately consistent with the first scan over regime I/II over the same amplitude range, suggesting that the amplitude could provide an approximate estimate of $E_{\mathrm{massless}}$ for any string configuration. The massive radiation is more subtle and complex; we have observed that we cannot satisfactorily compare $E_{\mathrm{massive}}$ between different string configurations with the same curvature or amplitude using our current modelling. However, we found that the ratio of the massive to massless radiation was approximated by a linear relationship $E_{{\mathrm{massive}}}/E_{{\mathrm{massless}}} \propto R_{\,\mathrm{Gaussian}}$ to one significant figure for $A \gtrsim 2$ for the scan over regimes II-IV. This may provide us with a method to probe $E_{\mathrm{massive}}$, even in highly nonlinear regimes. In the strongly nonlinear regime II, massive particle radiation made up approximately 50\% of the total signal, that is, an approximate equipartition between massless and massive modes for configurations sufficiently energetic to be well above the massive radiation threshold. 

We have found that the spectral tilt $q$ of the axion radiation, defined as $\mathcal{F}\left(E_{\,{\mathrm{massless}}}\right) \propto F^{-q}$ where $F$ is frequency and $\mathcal{F}$ denotes the Fourier transform, from the single-string, flat space configurations studied is bounded by $q \gtrsim 1$, where $q$ decreases as the amplitude $A$ of the initial travelling wave increases, i.e. the proportion of the axion signal in high energy modes increases. It is possible that this could shed light on current discrepancies in the spectral tilt observed from network simulations \cite{Buschmann_2022, Gorghetto:2021, Hindmarsh2021}. This is left to future work.

We have compared the radiation emitted from the travelling wave configurations with the sinusoidal configurations in \cite{Drew2019, Drew2023}. As previously determined, massive radiation cannot yet be meaningfully compared between the configurations. However, we have observed that $E_{\mathrm{massless}}$ emitted from different string configurations with similar amplitudes is equal to within approximately 50\%. This indicates that massless radiation from a string network or other burst configurations could be approximately parameterised by $A$, although the definition of amplitude in the context of a network is less clear.

Finally, we compared the results obtained for axion radiation from global strings to gravitational radiation from Nambu-Goto strings, motivated by the analogous forms for the power spectra obtained from Nambu-Goto and Kalb-Ramond models, detailed in \cite{Battye1993}. One important qualitative observation was that the massless spherical burst signals radiated from point at which the travelling waves collide are qualitatively similar to the isotropic GW signal predicted from kink-kink collisions in the Nambu-Goto model \cite{Binetruy2009}. Further work is needed to determine to what extent this analogy can be used to make quantitative claims about the local string GW signal using simulations of global string axion radiation. It is very interesting to note that it is possible to have highly nonlinear configurations (regime II) in which up to 50\% of the radiation can be radiated as massive particles. Although these are not the usual `cusps' or `kinks' on the boundary of nonlinearity in regime I, it may be that these highly nonlinear configurations may create bursts with both massive particles and gravitational waves, with the GW amplitude reduced by backreaction from linearized estimates. Additional future directions of study will be to generalise these observations to general network configurations, and perhaps to embed these into phenomenologcal network models, such as the velocity-one-scale (VOS) model \cite{Martins1996, Martins2002, Martins2019}.


\hspace{1cm}

\section*{Acknowledgements}

We are grateful for useful conversations with Ana Ach\'ucarro, Michalis Agathos, Pierre Auclair, Josu Aurrekoetxea, Richard Battye, David Benisty, José Ricardo Correia, Mark Hindmarsh, Tamanna Jain (also for her comments on an earlier draft), Eugene Lim, Javier Redondo, Ulrich Sperhake and Jenny Wagner, as well as the rest of the GRChombo team (http://www.grchombo.org/). We are grateful to Miren Radia for invaluable technical computing support. 

This publication is based upon work from COST Action COSMIC WISPers CA21106, supported by COST (European Cooperation in Science and Technology). We would like to thank the organisers of the Institute for Basic Science CTPU-CGA Workshop on Topological Defects in Daejeon for facilitating helpful and enlightening discussions, as well as the organisers of the workshops `Superconducting Cosmic Strings' at the Institut d'Astrophysique de Paris, and `Cosmic Topological Defects: Dynamics and Multimessenger Signatures' at the Lorentz Center, Leiden. We would like to acknowledge the support of the Intel Visualization team, notably the collaboration on in-situ visualization with Carson Brownlee and Jim Jeffers.

AD is supported by a Junior Research Fellowship (JRF) at Homerton College, University of Cambridge. Part of this work was undertaken whilst AD was supported by an EPSRC iCASE Studentship in partnership with Intel (EP/N509620/1, Voucher 16000206). TK was funded by a University of Cambridge Centre for Mathematical Sciences Summer Bursary under the Summer Undergraduate Research Project scheme, as well as from STFC Consolidated Grant ST/T00049X/1. EPSS acknowledges funding from STFC Consolidated Grant No. ST/P000673/1. AD and EPSS acknowledge project 2022.04048.PTDC, `Phi in the Sky,' from Funda\c{c}\~ao para a Ci\^encia e a Tecnologia (FCT), Portugal.

This work was undertaken on the Fawcett supercomputer at DAMTP funded by STFC Consolidated Grant ST/P000673/1, and the Cambridge CSD3 part of the STFC DiRAC HPC Facility (www.dirac.ac.uk). The DiRAC component of CSD3 was funded by BEIS capital funding via STFC Capital Grants ST/P002307/1 and ST/R002452/1 and STFC Operations Grant ST/R00689X/1. Scientific visualisation was performed using Kitware's open source visualisation application, ParaView, using the built-in ray tracing backend in conjunction with the Intel Rendering Toolkit and Intel OSPRay for enhanced rendering.

\bibliography{Paper1new}

\section*{Appendix A}\setcurrentname{Appendix A}\label{AppendixB}

\begin{figure*}
    \centering
    \includegraphics[width=0.48\textwidth]
    {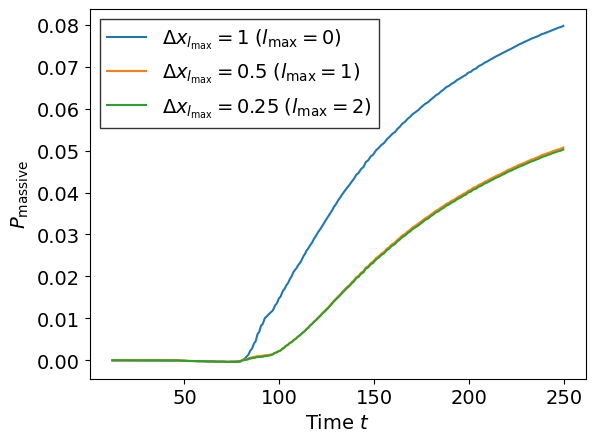}
    \includegraphics[width=0.48\textwidth]{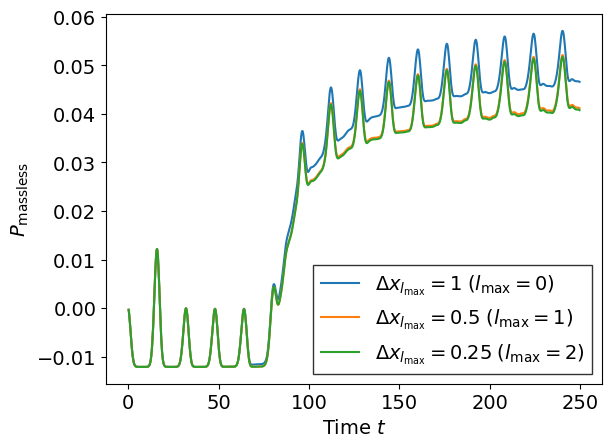} \\
    \includegraphics[width=0.48\textwidth]{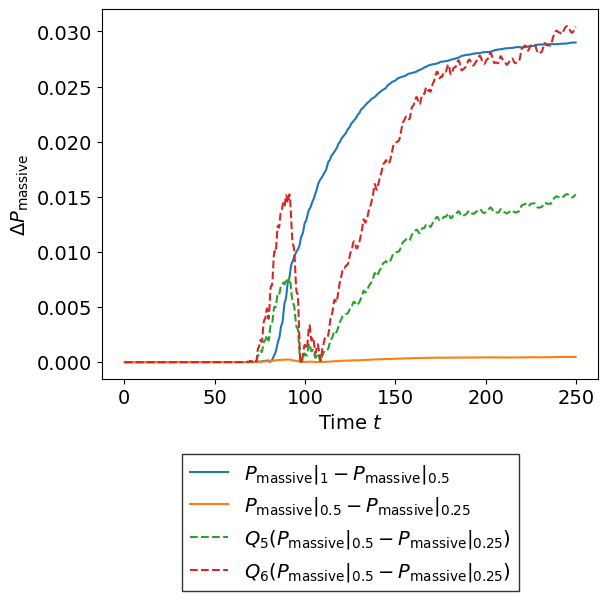}   
    \includegraphics[width=0.48\textwidth]{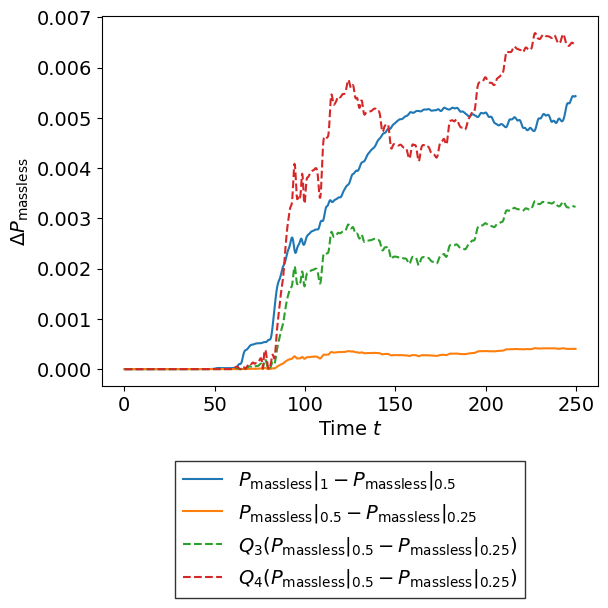}
    \caption{Absolute value (top) and convergence (bottom) of the energy emitted by massive radiation $P_\mathrm{massive}$ (left) and massless radiation $P_\mathrm{massless}$ (right) from a Gaussian/Gaussian travelling wave configuration with initial amplitude $A=1$ and $\sigma_{\mathrm d}=2$ using adaptive mesh refinement (test AMR in Table \ref{convergence_params}). The convergence plot shows the difference in the magnitude of $P_\mathrm{massive}$ and $P_\mathrm{massless}$ between different resolutions, with the higher resolution results also plotted rescaled according to $n$th-order convergence as indicated by the factors in the legends $Q_n$. In these plots, $P_\mathrm{massive}$ and $P_\mathrm{massless}$ are used as short-hand for $\int P_\mathrm{massive}\,\mathrm{d}t$ and $\int P_\mathrm{massless}\,\mathrm{d}t$.}
    \label{fig:AMRamp1convergence}
\end{figure*}

\begin{figure*}
    \centering
    \includegraphics[width=0.48\textwidth]{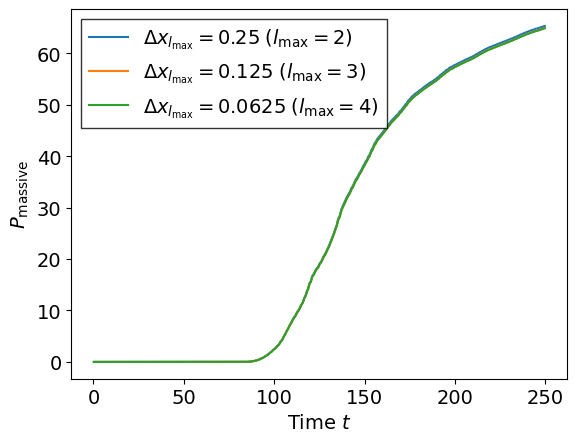}
    \includegraphics[width=0.48\textwidth]{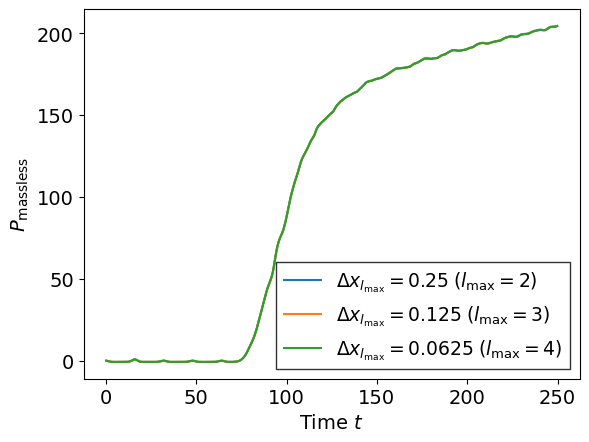}
    \includegraphics[width=0.48\textwidth]{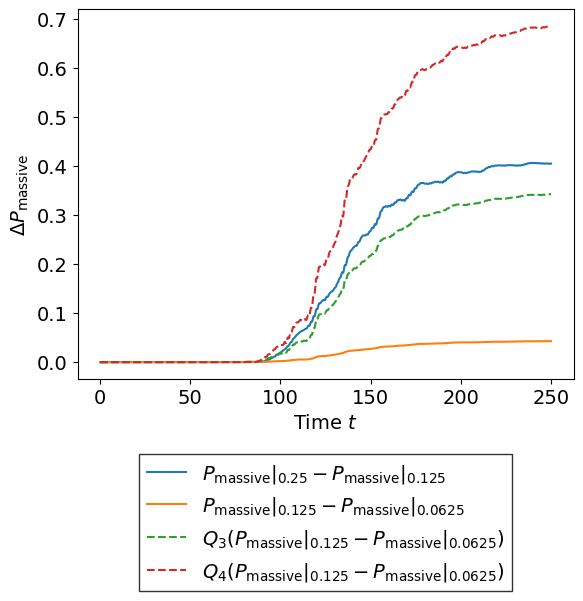}    
    \includegraphics[width=0.48\textwidth]{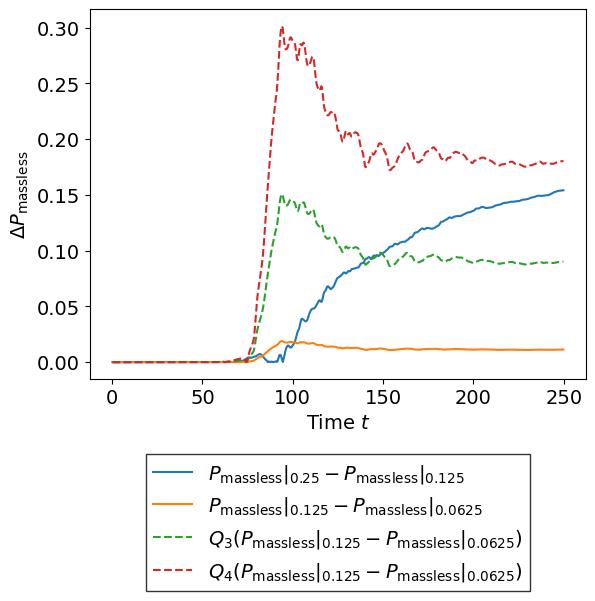}    
    \caption{Absolute value (top) and convergence (bottom) of the energy emitted by massive radiation $P_\mathrm{massive}$ (left) and massless radiation $P_\mathrm{massless}$ (right) from a Gaussian/Gaussian travelling wave configuration with initial amplitude $A=8$ and $\sigma_{\rm d}=2$ using adaptive mesh refinement (test AMR in Table \ref{convergence_params}). The convergence plot shows the difference in the magnitude of $P_\mathrm{massive}$ and $P_\mathrm{massless}$ between different resolutions, with the higher resolution results also plotted rescaled according to $n$th-order convergence as indicated by the factors in the legends $Q_n$. In these plots, $P_\mathrm{massive}$ and $P_\mathrm{massless}$ are used as short-hand for $\int P_\mathrm{massive}\,\mathrm{d}t$ and $\int P_\mathrm{massless}\,\mathrm{d}t$.}
    \label{fig:AMRamp8convergence}
\end{figure*}

\begin{figure*}
    \centering
    \includegraphics[width=0.48\textwidth]{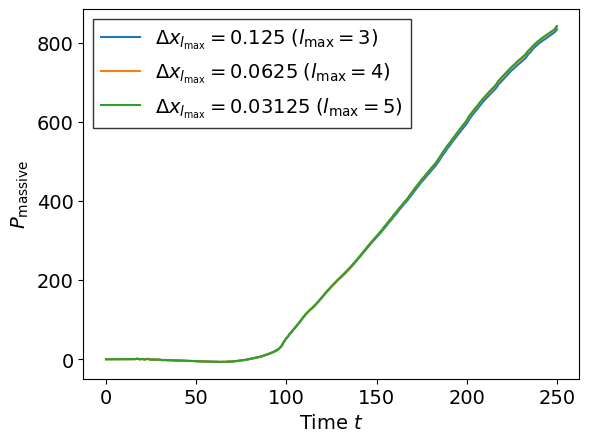}
    \includegraphics[width=0.48\textwidth]{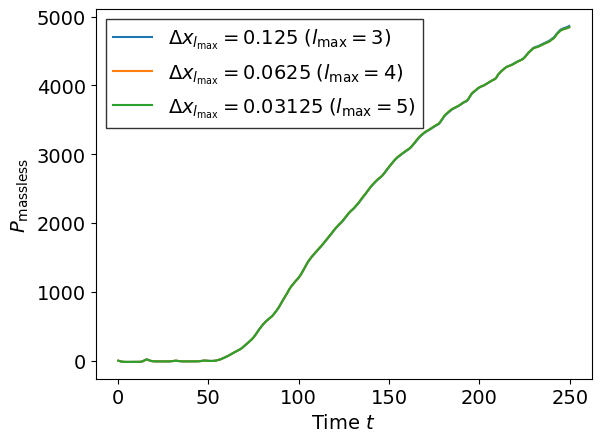}
    \includegraphics[width=0.48\textwidth]{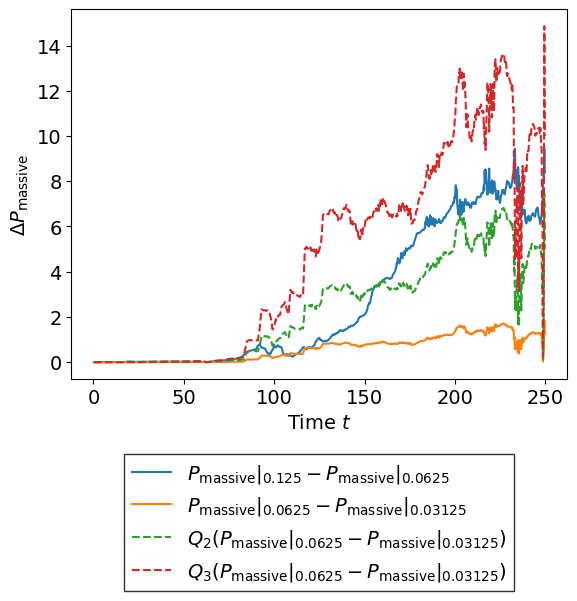}    
    \includegraphics[width=0.48\textwidth]{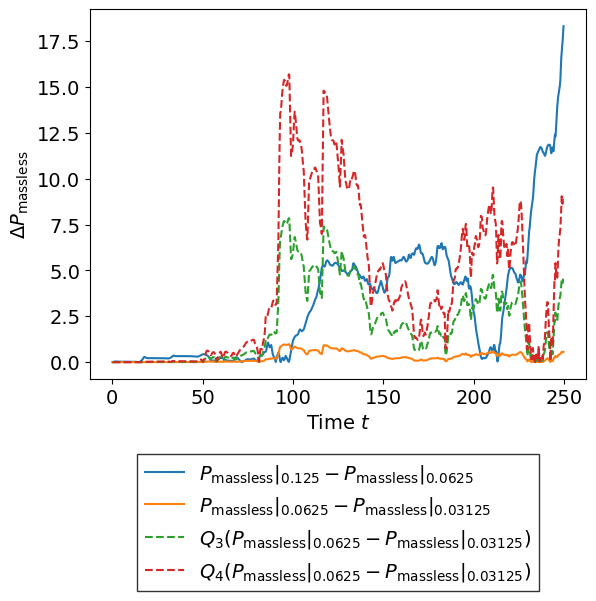}    
    \caption{Absolute value (top) and convergence (bottom) of the energy emitted by massive radiation $P_\mathrm{massive}$ (left) and massless radiation $P_\mathrm{massless}$ (right) from a Gaussian/Gaussian travelling wave configuration with initial amplitude $A=35$ and $\sigma_{\rm d}=2$ using adaptive mesh refinement (test AMR in Table \ref{convergence_params}). The convergence plot shows the difference in the magnitude of $P_\mathrm{massive}$ and $P_\mathrm{massless}$ between different resolutions, with the higher resolution results also plotted rescaled according to $n$th-order convergence as indicated by the factors in the legends $Q_n$. In these plots, $P_\mathrm{massive}$ and $P_\mathrm{massless}$ are used as short-hand for $\int P_\mathrm{massive}\,\mathrm{d}t$ and $\int P_\mathrm{massless}\,\mathrm{d}t$.}
    \label{fig:AMRamp35convergence}
\end{figure*}

\begin{figure*}
    \centering
    \includegraphics[width=0.48\textwidth]{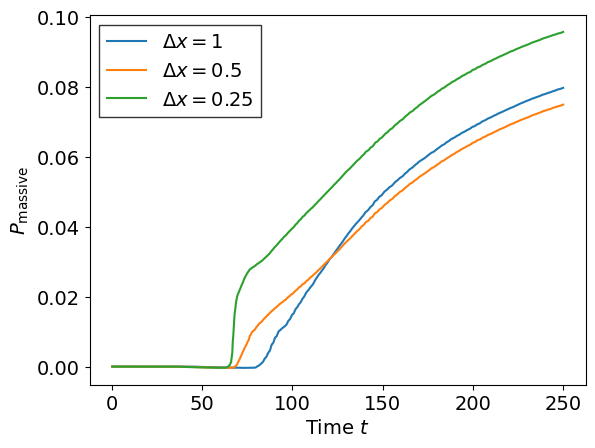}
    \includegraphics[width=0.48\textwidth]{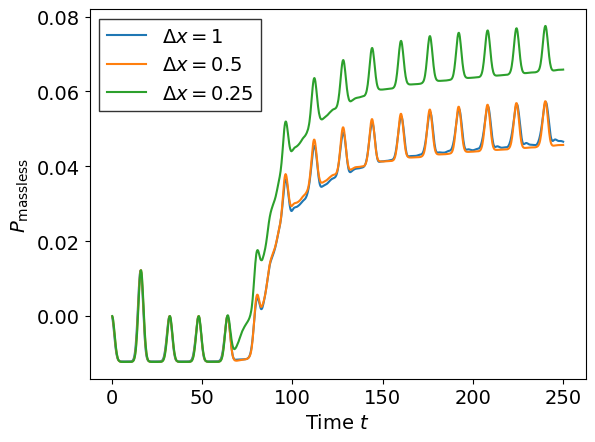} \\
    \includegraphics[width=0.48\textwidth]{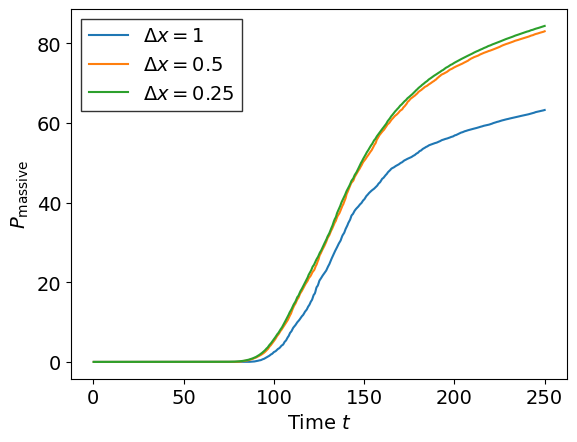}
    \includegraphics[width=0.48\textwidth]{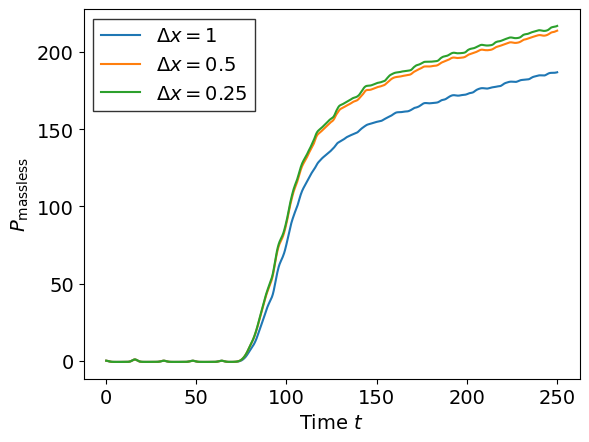}    
    \includegraphics[width=0.48\textwidth]{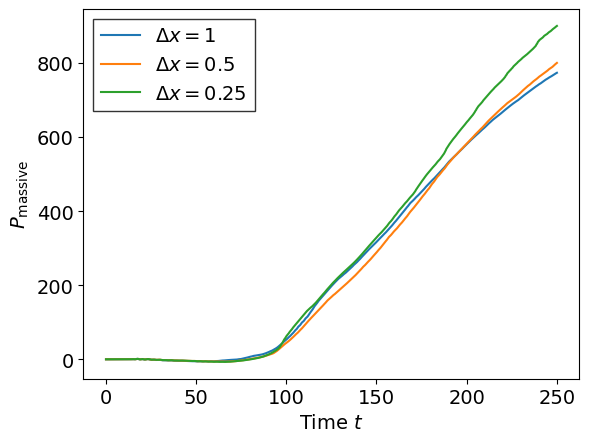}
    \includegraphics[width=0.48\textwidth]{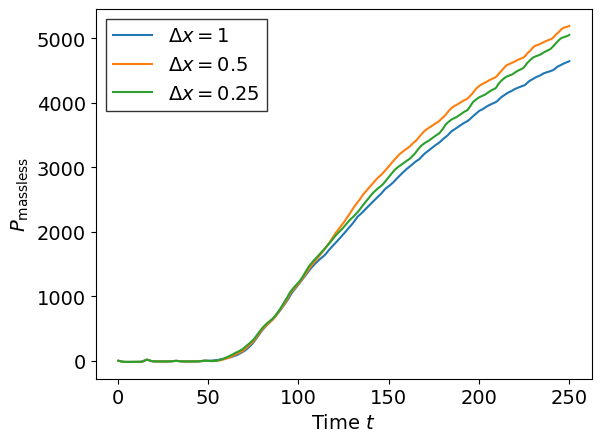} \\    
    \caption{Absolute value of the energy emitted by massive radiation $P_\mathrm{massive}$ (left) and massless radiation $P_\mathrm{massless}$ (right) from a Gaussian/Gaussian travelling wave configuration with initial amplitudes $A=1$ (top), $A=8$ (middle) and $A=35$ (bottom) and $\sigma_{\rm d}=2$ on a fixed grid (FG in Table \ref{convergence_params}). In these plots, $P_\mathrm{massive}$ and $P_\mathrm{massless}$ are used as short-hand for $\int P_\mathrm{massive}\,\mathrm{d}t$ and $\int P_\mathrm{massless}\,\mathrm{d}t$.}
    \label{fig:FGamp1convergence}
\end{figure*}

Here we present the parameters used for and the results of our convergence tests. The grid parameters are given in Table \ref{convergence_params}, and the convergence test plots in Figures \ref{fig:AMRamp1convergence}-\ref{fig:FGamp1convergence}.

\begin{table}[t]
    \centering
    \caption{Grid parameters for convergence tests. We perform convergence tests with AMR and compare to fixed grid results with different resolutions. For the fixed grid runs (\textit{FG}), the grid dimension $L$ remains constant and we change the base grid resolution $\Delta x_0$. For the AMR tests ($\textit{AMR}$), the maximum refinement level $l_{\rm max}$ is changed and $\Delta x_0$ remains constant. The base grid box resolution is given by $N^3$, with $(l_{\max}+1)$ total refinement levels including the coarsest base level, and grid spacings on the finest level given by $\Delta x_{l_{\rm max}}$.}
    \begin{ruledtabular}
    \begin{tabular}{cccccc}
        Test & $N$ & $l_{\max}$ & $L$ & $\Delta x_0$ & $\Delta x_{l_{\rm max}}$  \\
                 \hline
        \textit{FG} & $512$ & - & 512 & 1 & - \\
         & $1024$ & - & 512 & 0.5 & - \\
         & $2048$& - & 512 & 0.25 & - \\
        \hline
        \textit{AMR} & $512$ & 0 & 512 & 1 & 1 \\
        & $512$ & 1 & 512 & 1 & 0.5 \\
        & $512$ & 2 & 512 & 1 & 0.25 \\
        & $512$ & 3 & 512 & 1 & 0.125 \\
         & $512$ & 4 & 512 & 1 & 0.0625  \\
        & $512$ & 5 & 512 & 1 & 0.03125  \\        & $512$ & 6 & 512 & 1 & 0.015625  \\ 

    \end{tabular}
    \end{ruledtabular}
    \label{convergence_params}
\end{table}

\section*{Appendix B}\setcurrentname{Appendix B}\label{AppendixC}

Here we outline the analogy between gravitational radiation from local strings and axion radiation from global strings using the Nambu-Goto and Kalb-Ramond models. The expressions below are valid for periodic configurations with periodicity $L$, but can also be derived for general configurations.

It has been shown \cite{Sakellariadou:1991sd, Sakellariadou1990} that the power per unit length radiated from an infinitely long periodic global string lying along the $z$-direction via massless radiation can be written in terms of the Fourier transform $\Tilde{J}^{\mu\nu}$ of the source distribution $J^{\mu\nu}$. This is given by
\begin{align}\label{longspectrum}
\frac{dP}{dz} = 2\pi\sum^\infty _{n=1}&\omega_n \sum_{|\kappa_m|< \omega_n} \\ &\int^{2\pi}_0 {\mathrm{d}\theta \, \Tilde{J}^{\mu\nu *}(\omega_n,\mathbf{k}^\perp, \kappa_m})\,\Tilde{J}_{\mu\nu}(\omega_n,\mathbf{k}^\perp, \kappa_m)\,, \nonumber
\end{align} 
where $\omega_n = 2\pi n/L$, $\kappa_m = 2\pi m / \alpha L$, where $\alpha$ is the length of the periodic configuration relative to $L$, $\mathbf{k}^\perp = |\mathbf{k}^\perp|(\cos\theta, \sin\theta)$ and $|\mathbf{k}^\perp| = \sqrt{\omega_n^2 - \kappa_m^2}$.

We can write an analogous expression for the gravitational radiation from a local string in an equivalent configuration, given by
\begin{align}\label{longspectrumGW}
\frac{dP}{dz} = 2G\sum^\infty _{n=1}\omega_n & \sum_{|\kappa_m|< \omega_n} \\ \int^{2\pi}_0 \mathrm{d}\theta & \, \left(\Tilde{T}^{\mu\nu *}(\omega_n,\mathbf{k}^\perp, \kappa_m)\,\Tilde{T}_{\mu\nu}(\omega_n,\mathbf{k}^\perp, \kappa_m) \right. 
\nonumber \\ & \left. - \frac{1}{2}\left|\Tilde{T}_\lambda^\lambda(\omega_n,\mathbf{k}^\perp, \kappa_m) \right|^2\,\right), \nonumber
\end{align}
where $T^{\mu\nu}$ is the energy-momentum tensor of the string. See \cite{Battye1993} for further details.


\end{document}